\documentclass[reprint,twocolumn,superscriptaddress,secnumarabic,amssymb,nobibnotes,prb]{revtex4-2}
\usepackage{graphicx}
\usepackage{dcolumn}
\usepackage{bm}
\usepackage{xcolor}
\usepackage[export]{adjustbox}
\usepackage{float}
\usepackage{tabularx}
\usepackage{multirow}
\usepackage{amsmath}
\usepackage{braket}
\usepackage{array}
\usepackage{ulem}
\usepackage{comment} 
\usepackage[utf8]{inputenc}

\newcolumntype{C}[1]{>{\centering\let\newline\\\arraybackslash\hspace{0pt}}m{#1}}

\begin{document}
	
	
	\title{The Kondo effect in ferromagnetic quantum critical CeRh$_6$Ge$_4$}

	\author{Martin~Sundermann}
	\affiliation{Max Planck Institute for Chemical Physics of Solids, N{\"o}thnitzer Stra{\ss}e 40, 01187 Dresden, Germany}
	\affiliation{PETRA III, Deutsches Elektronen-Synchrotron DESY, Notkestra{\ss}e 85, 22607 Hamburg, Germany}
	
	\author{Joe~D.~Thompson}
	\affiliation{Los Alamos National Laboratory, Los Alamos, New Mexico 87545, USA}
	
	\author{Eric~D.~Bauer}
	\affiliation{Los Alamos National Laboratory, Los Alamos, New Mexico 87545, USA}
	
	\author{Chun-Fu~Chang}
	\affiliation{Max Planck Institute for Chemical Physics of Solids, N{\"o}thnitzer Stra{\ss}e 40, 01187 Dresden, Germany}
	
	\author{Sheng-Huai~Chen}
	\affiliation{Max Planck Institute for Chemical Physics of Solids, N{\"o}thnitzer Stra{\ss}e 40, 01187 Dresden, Germany}
	
	\author{Chang-Yang~Kuo}
	\affiliation{Department of Electrophysics, National Yang Ming Chiao Tung University, Hsinchu, 30010, Taiwan  }
	\affiliation{National Synchrotron Radiation Research Center, 101 Hsin-Ann Road, Hsinchu 300092, Taiwan  }
		
	\author{Liu~Hao~Tjeng}
	\affiliation{Max Planck Institute for Chemical Physics of Solids, N{\"o}thnitzer Stra{\ss}e 40, 01187 Dresden, Germany}
	
	\author{Gertrud~Zwicknagl}
	\affiliation{Technische Universität Braunschweig, 38106 Braunschweig, Germany}
	\affiliation{Max Planck Institute for Chemical Physics of Solids, N{\"o}thnitzer Stra{\ss}e 40, 01187 Dresden, Germany}
	
	\author{Andrea~Severing}
	\affiliation{Max Planck Institute for Chemical Physics of Solids, N{\"o}thnitzer Stra{\ss}e 40, 01187 Dresden, Germany}
	
	\date{\today}

	\begin{abstract}
	The mechanism of a pressure-induced quantum critical point in the heavy fermion ferromagnet CeRh$_6$Ge$_4$ has attracted interest, as ferromagnetic quantum criticality in a clean itinerant Ce compound is typically avoided. The localized versus itinerant character of the 4\textit{f} electrons is a key aspect for understanding this behavior. We investigated the electronic structure of the 4\textit{f} shell in CeRh$_6$Ge$_4$ using core-level photoelectron and x-ray absorption spectroscopy, demonstrating the hybridization of Ce 4\textit{f} with the conduction electrons. Linearly polarized x-ray absorption reveals a temperature-dependent linear dichroism consistent with the crystal-electric-field (CEF) sequence as inferred from the static susceptibility. This dichroism cannot be described by an ionic full-multiplet model alone, but is reproduced by including the Kondo effect within a single-impurity Anderson model in the non crossing approximation (SIAM/NCA). The Kondo effect mixes higher lying crystal-field states into a resulting multiorbital ground state with 4\textit{f} occupancy \textit{n}$_f$\,$\sim$\,0.9. Deviations at low temperatures between the measured linear dichroism and calculated  dichroism  suggest an orbital-dependent Kondo effect. A scenario in which there is a multiorbital ground state and orbital-dependent Kondo hybridization should be a starting point for a model of pressure-induced criticality in CeRh$_6$Ge$_4$.	
			
	\end{abstract} 
	
	\maketitle
	
	\section{Introduction}
	Ferromagnetic (FM) quantum critical points (QCP) in clean itinerant Kondo materials are usually avoided, preempted either by first-order phase transitions or by the formation of inhomogeneous magnetic phases\,\cite{Belitz1999,Chubukov2004,Belitz2005,Conduit2009,Brando2016,Kirkpatrick2020}. 	The metallic compound YbNi$_4$P$_2$ undergoes a FM quantum phase transition upon substitution of As on the P sites, however, such  substitution also introduces disorder\,\cite{Steppke2013}. In contrast, stoichiometric CeRh$_6$Ge$_4$ orders ferromagnetically at  $T_{\rm C}$\,=\,2.5\,K\,\cite{Vosswinkel2012,Matsuoka2015} out of a strongly correlated paramagnetic state and exhibits a FM\,QCP under external pressure at $p_c$\,=\,0.8GPa. Moreover, at its QCP the material displays strange-metal behavior -- a logarithmic divergence of specific heat divided by temperature and linear-in-temperature resistivity\,\cite{Kotegawa2019,Shen2020}, which commonly occurs in quantum-critical antiferromagnets\,\cite{Paschen2020}. Theoretically a FM\,QCP in a \textit{clean} material can arise in itinerant ferromagnets if the system is quasi-one-dimensional\,\cite{Kirkpatrick2020,Komijani2018} or if there is strong antisymmetric spin-orbit coupling due to broken inversion symmetry \cite{Kirkpatrick2020,Shin2024,Yamamoto2010}.  If, however, magnetic electrons are localized in a Kondo lattice criticality could arise from breakdown of the Kondo effect at the QCP, producing a change in Fermi volume from small to large for coherent  quasi particles\,\cite{Shen2020,Komijani2018}. A change in Fermi surface would also be consistent with over-damped phase fluctuations of hybridization that could account for both ferromagnetic criticality and strange metal behaviors\,\cite{Zhan2025}. Each of these scenarios has been invoked to account for criticality in  CeRh$_6$Ge$_4$\,\cite{Shen2020,Zhan2025,Shin2024,Wang2021,Wu2021,Thomas2024,Itokazu2025}, which emphasizes the conundrum it poses. Distinguishing characters of these scenarios is the nature of the 4\textit{f} electron and the role played by the Kondo effect. Clarifying these characters is a necessary step to resolving the conundrum. 
	
	CeRh$_6$Ge$_4$ crystallizes in the hexagonal LiCo$_6$P$_4$-type structure (P\underline{6}m2) without an inversion center\,\cite{Vosswinkel2012}.  The structure consists of Ce-Rh1-Ge1 and Rh2-Ge2 layers, with triangular lattices of Ce stacked along the $c$-axis, giving rise to chains of Ce along this direction. Ce occupies the 1a sites with $D_{\rm 3h}$ point symmetry, in which Ce ions experience a hexagonal (\underline{6}m2) crystal-electric field (CEF) that splits the  $J$\,=\,5/2 multiplet into three Kramers doublets, each corresponding to a pure $j_z$ state ($|\pm1/2 \rangle$, $|\pm3/2 \rangle$, $|\pm5/2 \rangle$). 
	
	The magnetic properties of CeRh$_6$Ge$_4$ are very anisotropic with the crystallographic \textit{c} direction being the hard axis, thus restricting the ordered moments to the \textit{ab} plane. The effective moment $\mu_{\rm eff}$ amounts to 2.35$\mu_{\rm B}$/Ce, which is close to the full Ce$^{\rm 3+}$ moment, and the ordered moments as determined from magnetization\,\cite{Shen2020}, muon spin resonance ($\mu$SR)\,\cite{Shu2021} and nuclear quadrupole resonance (NQR)\,\cite{Yamamoto2025} are of the order of 0.2-0.3\,$\mu_{\rm B}$/Ce. This is much smaller than the in-plane expectation value of 1.28\,$\mu_{\rm B}$/Ce for the $|\pm1/2 \rangle$ CEF ground state. Because $|\pm1/2 \rangle$ is the only state that corresponds to a finite in-plane moment, it is deduced to be ground state. Fits to the anisotropy of the static susceptibility yield the $|\pm3/2 \rangle$ CEF state 5.8\,meV and the $|\pm5/2 \rangle$ CEF state at 22.1\,meV\,\cite{Shu2021}.
	
	Angle-dependent quantum oscillation measurements at ambient pressure indicate that CeRh$_6$Ge$_4$ is a \textit{clean} material and are more consistent with band-structure calculations that assume that the 4\textit{f} electrons do not contribute to the Fermi surface at ambient pressure. This suggests that CeRh$_6$Ge$_4$ does not form a coherent Fermi liquid and that Kondo-breakdown criticality may be relevant for a description of quantum criticality in CeRh$_6$Ge$_4$\,\cite{Wang2021}.	
	
	Alternatively, the small ordered moment\,\cite{Shen2020,Shu2021,Yamamoto2025}, large Sommerfeld coefficient ($\sim$250\,mJ/molK$^2$) deep in the ordered state and the small fraction of entropy (about 20\% of $R\ln 2$) released at the ordering transition\,\cite{Matsuoka2015}, are more indicative of a local Fermi liquid and a substantial Kondo effect. This is consistent with inelastic neutron scattering results, which reveal a broad quasielastic linewidth of 3.3\,meV ($\approx$40\,K) and a strongly broadened inelastic response\,\cite{Shu2021}, as well as with angle-resolved photoelectron spectroscopy (ARPES) that reveals dispersive 4\textit{f} bands and anisotropic hybridization between 4\textit{f} and conduction electrons well above \textit{T}$_{\rm C}$ \cite{Wu2021}. In fact, the Kondo-derived peak observed in ARPES persists to temperatures well above both the Kondo temperature \textit{T}$_{\rm K}$\,$\sim$\,20\,K estimated from entropy\,\cite{Matsuoka2015}, and the coherence temperature estimated from resistivity, \textit{T}$^{*} \sim 80$ K \cite{Wu2021}.  Thermopower measurements question a Kondo-breakdown scenario with fully localized moments\,\cite{Thomas2024}. Instead, the results of these measurements\,\cite{Thomas2024} argue that the conditions for quantum criticality of itinerant ferromagnetism are fulfilled due to the formation of a quasiquartet ground state arising from the intermixing of CEF states in the presence of the Kondo effect. They further propose that orbital selective hybridization, supported by APRES\,\cite{Wu2021}, may reconcile the discrepancies among the various experimental findings.
	
	From Ce\,3\textit{d} core-level photoelectron (PES) and \textit{M}$_{4,5}$ edge polarized x-ray absorption (XAS) spectroscopies, we address two outstanding questions posed by CeRh$_6$Ge$_4$: what is the nature of its 4\textit{f} electrons and what is an appropriate description of its crystal-field excitations? Satellites in the core-level PES  provide insight into the presence of Kondo coupling and the associated  configurational mixing. At low temperatures, linearly polarized XAS gives access to the orbital occupation of the 4\textit{f} shell ground state and allows us to probe possible mixing of excited CEF states into the ground state induced by the Kondo effect. 
		
	\section{Method: experiment and analysis}
    CeRh$_6$Ge$_4$ single crystals were grown with the Bi-flux technique as described by Vosswinkel et al.\,\cite{Vosswinkel2012}. X-ray diffraction pattern confirmed the \textit{P6m2} hexagonal structure with lattice constants  \textit{a}\,=\,7.15\,\AA\, and \textit{c}\,=\,3.85\,\AA.  Magnetic susceptibility and magnetization measurements performed in an MPMS3 magnetometer (Quantum Design) gave a ferromagnetic transition at 2.5\,K and agreement with previously reported magnitude and anisotropy of magnetic properties of CeRh$_6$Ge$_4$\,\cite{Matsuoka2015,Shen2020,Shu2021}.
	
	PES and XAS measurements were performed at the NSRRC-MPI TPS-45A1 endstation at the Taiwan Photon Source (TPS) with a beam focus of V\,x\,H = 25\,$\mu$m\,x\,30\,$\mu$m. Crystals of CeRh$_6$Ge$_4$ were cleaved \textit{in situ} under ultra high vacuum (UHV) conditions of p$<$10$^{-10}$\,mbar to ensure a clean surface. 

	PES was measured at 200\,K with horizontally polarized light at a fixed photon energy of 1260\,eV. The photoelectrons were detected with an MBS A-1 analyzer with a lens-4 detector, placed at an angle of 120$^{\circ}$ between photon and electron wave vectors in the horizontal plane. The energy resolution was 175\,meV as measured on a Pt reference with the Fermi energy at 1256.3\,eV. A polynomial correction was applied to straigthen the spectrum prior to subtracting a Shirley-type background.

	XAS measurements at the Ce \textit{M}$_{4,5}$ edges were performed in the total-electron-yield (TEY) mode with linearly polarized light and an energy resolution of about 100\,meV. All spectra were normalized to the incident photon flux, calibrated in energy to a CuO reference, and background subtracted using the pre-edge intensity. The single crystal was aligned such that the electric field vector $\vec{E}$ was either perpendicular to the crystallographic \textit{c} axis ($\vec{E}$\,$\perp$\,\textit{c}) or approximately parallel to it ($\vec{E}$\,$\parallel$\,$\tilde{c}$, with $\tilde{c}$ 20$^\circ$\,off\,\textit{c}, due to geometric restrictions). The presented $\vec{E}$\,$\parallel$\,\textit{c} spectra are corrected for this offset as described in the Appendix. 

	Core-level PES spectra were analyzed using full-multiplet (\textit{fm}) configuration-interaction (CI) calculations performed with the Quanty code \cite{Haverkort2016}. The model includes a single, energetically degenerate ligand shell hybridizing equally with all 4\textit{f} states, preserving spherical symmetry \cite{Imer1987}, and was restricted to the 4\textit{f}$^0$, 4\textit{f}$^1$\underline{\textit{L}}, and 4\textit{f}$^2$\underline{\underline{\textit{L}}} configurations. Here \textit{\underline{L}} and \underline{\underline{\textit{L}}} denote the number of ligand holes.
	Atomic spin–orbit coupling values from the Cowan code\,\cite{CowanBook} were used for both 4\textit{f} and 3\textit{d} shells, while Coulomb interactions, starting with the Cowan values, were reduced to 60\% (4\textit{f}–4\textit{f}) and 80\% (3\textit{d}–4\textit{f}) to account for screening effects within the solid\,\cite{Tanaka1994,Groot2008}. The spectral lines are broadened using a combination of Gaussian (FWHM\,=\,1.1\,eV) and Lorentzian (FWHM\,=\,1.4\,eV) functions to account for instrumental resolution and lifetime broadening. Additionally, a Mahan function is applied to capture the asymmetry of the line shapes (cut-off parameter $\xi_M$\,=\,4 eV, asymmetry factor $\alpha_M$\,=\,0.15) (see, e.g., in\,\cite{Strigari2015}).  Best agreement with experiment was obtained for the CI parameters $\epsilon_f$\,=\,2.8 eV (4\textit{f} binding energy), \textit{U}$_{ff}$\,=\,9.7\,eV (Coulomb repulsion in the 4\textit{f} shell), \textit{V}$_{\rm eff}$\,=\,0.29\,eV (effective hybridization), and \textit{U}$_{fc}$\,=\,10\,eV (core-hole potential). The ground state configurational mixing is reliably determined because the four CI parameters are constrained from the energy splittings and intensity ratios\,\cite{Gunnarsson1983}.
	
	For the analysis of the \textit{M}$_{4,5}$ XAS data of Ce$^{3+}$ configuration (3\textit{d}$^{10}$4\textit{f}$^1$\,$\rightarrow$\,3\textit{d}$^{9}$4\textit{f}$^2$), ionic \textit{fm} calculations were performed  with the atomic Cowan values of the atomic 4\textit{f} (3\textit{d}) spin-orbit interaction reduced to 95\% (97\%) and the atomic 4\textit{f}-4\textit{f} (3\textit{d}-4\textit{f}) Coulomb repulsion lowered to 60\% (80\%).
	The isotropic spectra are constructed from the polarized spectra as I$_{\rm iso}$\,=\,1/3$\cdot$I($\vec{E}$\,$\parallel$\,\textit{c})\,+\,2/3$\cdot$I($\vec{E}$\,$\perp$\,\textit{c}).
	As discussed in the Appendix, \textit{fm} calculations of these isotropic spectra agree well with experimental results.
	
	The linear dichroism, \textit{LD}\,=\,I(${E \parallel c}$)\,-\,I(${E \perp c}$), was calculated for the three pure \textit{j}$_z$ CEF eigenstates of the Ce$^{3+}$ \textit{J}\,=\,5/2 ground state multiplet in $D_{\rm 3h}$ point symmetry. At low temperatures this reflects the ground state orbital occupation, while its temperature dependence is due to thermal occupation of excited CEF states\,\cite{Hansmann2008}. 	Deviations from \textit{fm} calculations are likely due to Kondo induced mixing of CEF states\,\cite{Sundermann2015,Wissgott2016,Amorese2023,Christovam2024}. Notably, the Ce$^{4+}$\,4\textit{f}$^0$ configuration with \textit{J}\,=\,0 does not contribute to the \textit{LD}.

	\section{Results and Discussion}
\begin{figure}[t]
		\begin{center}
				\includegraphics[width=0.97\columnwidth]{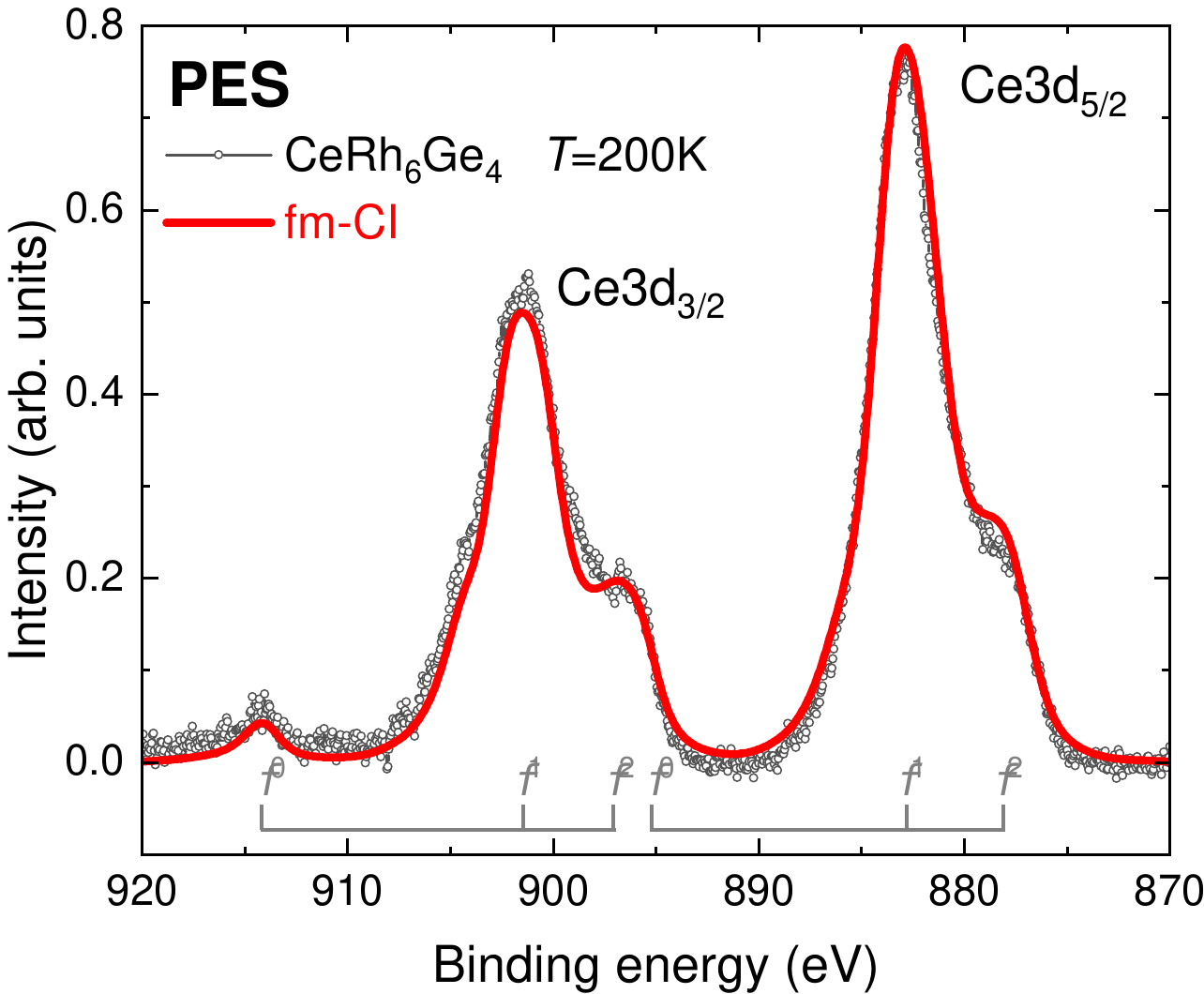}
			\end{center}
		\caption{Ce\,3d core level PES data of CeRh$_6$Ge$_4$ (black circles) after background correction, plus the result of a full-multiplet configuration-interaction  (FM-CI) simulation (red line). The gray ruler at the bottom indicates the energy positions of the I(f$^n$) spectral weights.}
		\label{PES}
	\end{figure}

	In Fig.\,\ref{PES}, the Ce\,3\textit{d} core-level PES measurements of CeRh$_6$Ge$_4$ are shown. The dominant spectral features are assigned to Ce$^{3+}$\,4\textit{f}$^1$ in the initial state. In addition, weaker satellites corresponding to the 4\textit{f}$^0$ and 4\textit{f}$^2$ configurations are observed, indicating a mixed-valence ground state of Ce. For example, the red arrow points out the 4\textit{f}$^0$ spectral weight of the Ce\,3\textit{d}$_{3/2}$ emission line and the gray rulers at the bottom of the figure indicate the energy positions of all 4\textit{f}$^n$ spectral weights. These results provide direct evidence for significant hybridization between the Ce\,4\textit{f} states and the conduction electrons in CeRh$_6$Ge$_4$ and reinforce conclusions from ARPES\,\cite{Wu2021}.
	
	To quantitatively assess the degree of hybridization, a \textit{fm}-CI analysis, as described in Sec.\,II, was performed. The optimized CI parameters, summarized in Sec.\,II, yield ground-state configuration weights of 8\% 4\textit{f}$^{0}$, 89\% 4\textit{f}$^{1}$, and 3\% 4\textit{f}$^{2}$, i.e., a total occupation of the 4\textit{f} shell n$_{f}^{\rm tot}$\,=\,0.95 at 200\,K. Comparison with other cerium compounds analyzed within the same simplified \textit{fm}-CI model suggests that CeRh$_6$Ge$_4$ is similar to CeRh$_3$B$_2$ in terms of hybridization strength and configurational mixing\,\cite{Sundermann2016,Dhar1981}, although the effects are less pronounced in CeRh$_6$Ge$_4$.
		
 	\begin{figure}[t]
		\begin{center}
			\includegraphics[width=0.99\columnwidth]{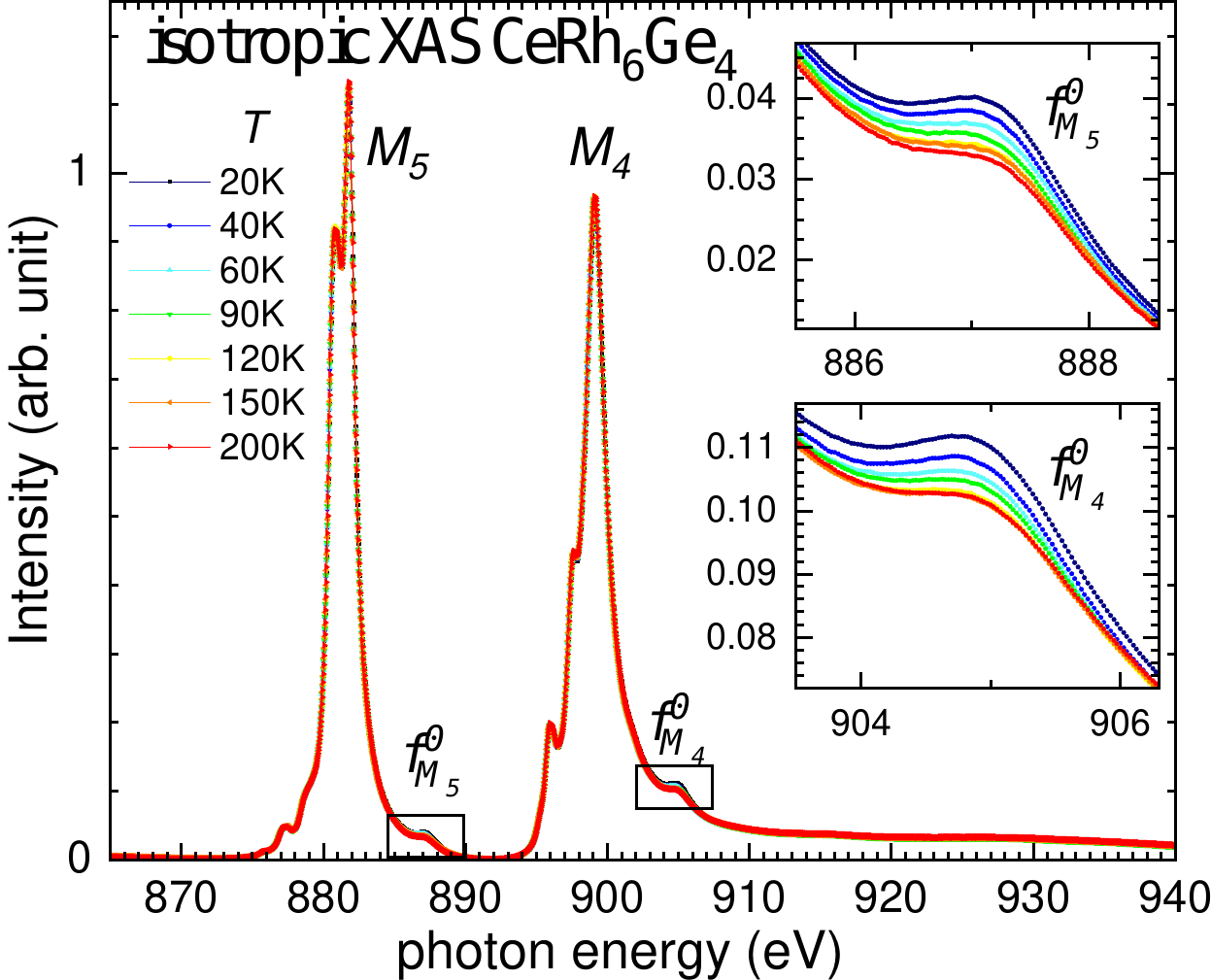}
		\end{center}
		\caption{Isotropic XAS spectra, I$_{\rm iso}$\,=\,1/3$\cdot$(I(${E \parallel c}$)\,+\,2$\cdot$I(${E \perp c}$)), of CeRh$_6$Ge$_4$ at the Ce\,\textit{M}$_{4,5}$ edges for several temperatures \textit{T} between 20 and 200\,K. Insets: enlarged view of 4\textit{f}$^0$ satellites at the high energy tail of the \textit{M}$_5$ and \textit{M}$_4$ edges.
		}
		\label{ISO}
	\end{figure}
	XAS also is sensitive to configuration-interaction effects when the ground state is not purely of 4\textit{f}$^{1}$ character. Hybridization between the Ce\,4\textit{f} and conduction-electrons induces a finite 4\textit{f}$^{0}$ component in the ground (initial) state, giving rise to satellites on the high-energy side of the main Ce$^{\rm 3+}$ (4\textit{f}$^{1}$) absorption peaks. If these 4\textit{f}$^{0}$ satellite features originate from Kondo screening, their spectral weight is expected to decrease with increasing temperature.
	
	In Fig.\,\ref{ISO}, the isotropic XAS spectra of CeRh$_6$Ge$_4$ are shown for several temperatures between 20 and 200\,K. These isotropic spectra are constructed from the measured liner polarized data as described in section II. The 4\textit{f}$^0$ satellites are highlighted by rectangular frames and are displayed on an enlarged scale in the insets. Decreasing spectral weights with increasing temperature are observed, indicative of the Kondo effect - which is consistent with the core-level PES analysis above. Notably, also in XAS the 4\textit{f}$^0$ spectral weights remain pronounced at 200\,K.
	  
	We now turn to the linearly polarized data. The linearly polarized XAS data of CeRh$_6$Ge$_4$, measured with $\vec{E}$\,$\perp$\,\textit{c} (blue) and $\vec{E}$\,$\parallel$\,\textit{c} (red), are shown for all temperatures in the Appendix. 
	Figure\,\ref{LD}\,(a) presents the linearly polarized XAS spectra of CeRh$_6$Ge$_4$ at 20\,K only, together with the experimental linear dichroism \textit{LD}$_{\rm exp}$ (green). A pronounced polarization dependence is observed. 
	
	We first compare the experimental data in Fig.\,\ref{LD}\,(a) qualitatively with simulated polarized XAS spectra for the pure $|\pm j_z \rangle$ states. Figure\,\ref{LD}\,(b) shows the simulation for the \textit{M}$_5$ edge, scaled to the experimental data. Both the $|\pm1/2 \rangle$ and $|\pm3/2 \rangle$ states reproduce qualitatively the  polarization dependence. However, a direct comparison of the experimental linear dichroism, \textit{LD}$_{\rm exp}$, with the simulated dichroism \textit{LD}$_{j_z}$, shown in Fig.\,\ref{LD}\,(c), reveals that \textit{LD}$_{1/2}$ significantly overestimates while \textit{LD}$_{3/2}$ underestimates the dichroism. 
	
	We recall that the \textit{LD}$_{1/2}$ state is expected to form the ground state and that, keeping in mind thermal population of the CEF levels, its occupation at 20\,K should be approximately 95\%, while the $|\pm3/2 \rangle$ state at 5.8\,meV should contribute about 5\% (see dashed lines in Fig.\,\ref{Tdep}\,(d)). It is evident that the CEF scheme considered here, including the thermal population of excited states, cannot reproduce the experimentally observed dichroism at 20\,K. However, the low-temperature dichroism can be reproduced either by assuming a $|\pm 1/2\rangle$  ground state with a $|\pm 5/2\rangle$ excited state close in energy and partially populated, or by considering a  $|\pm 3/2\rangle$ ground state with partial thermal population of a $|\pm 1/2\rangle$ excited state, but both scenarios are inconsistent with the anisotropy of the static magnetic susceptibility and, in the latter case, the experimentally observed in-plane alignment of the ordered moments.

		\begin{figure}[t]
		\begin{center}
			\includegraphics[width=0.95\columnwidth]{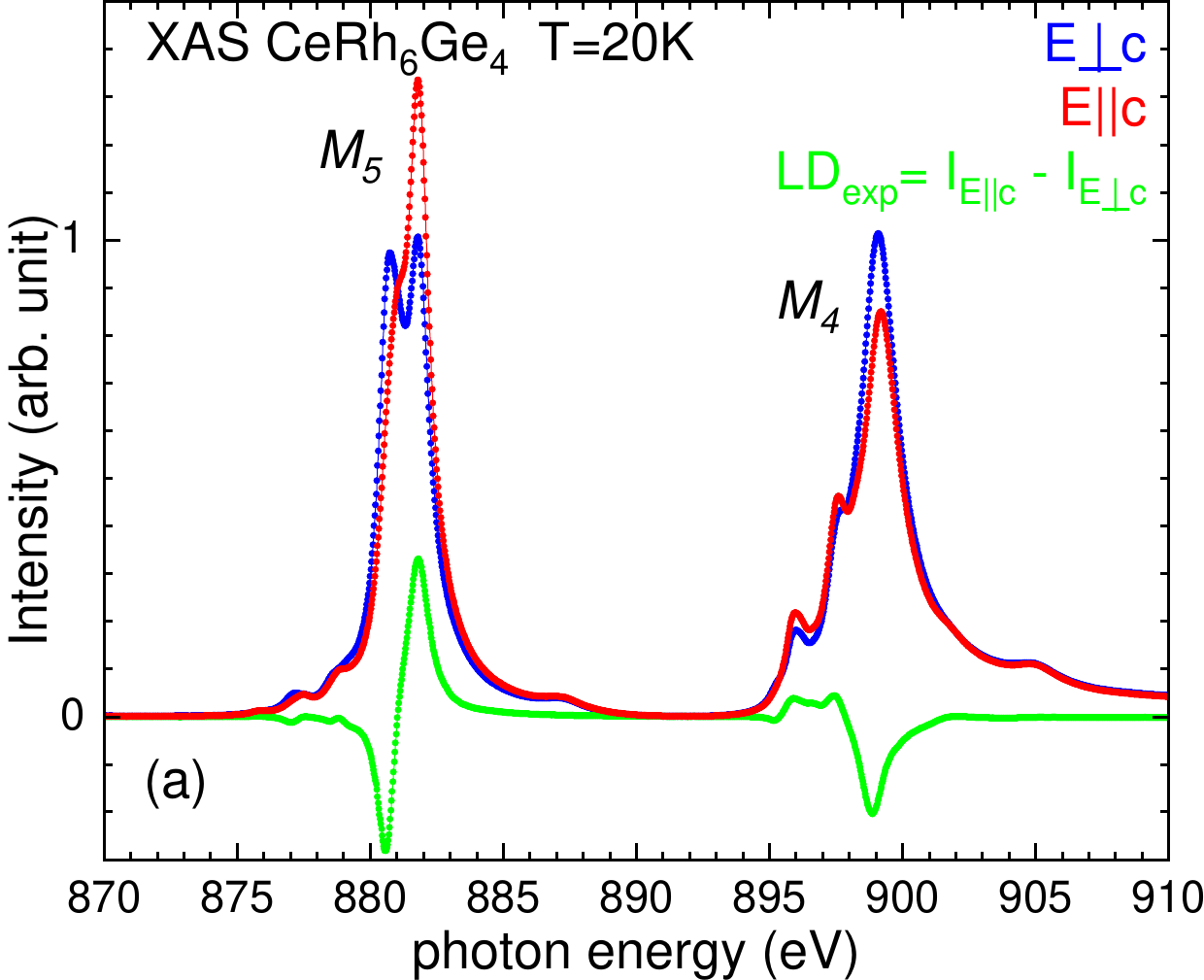}
			\includegraphics[width=0.95\columnwidth]{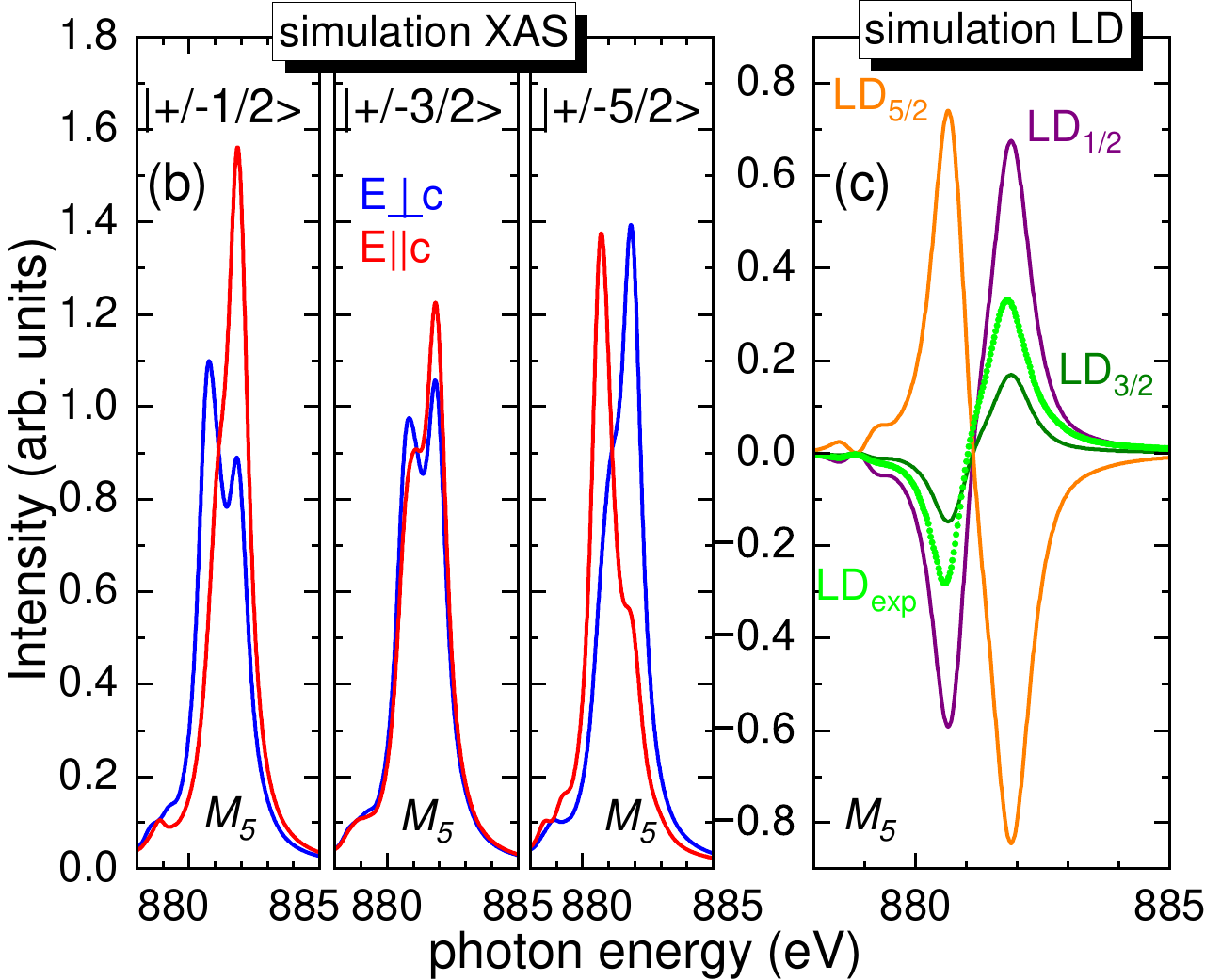}
		\end{center}
		\caption{(a) Linear polarized XAS spectra, I(${E \parallel c}$) (red) and I(${E \perp c}$) (blue), of CeRh$_6$Ge$_4$ at the Ce\,\textit{M}$_{4,5}$ edges at \textit{T}\,=\,20K and the corresponding linear dichroism \textit{LD}$_{\rm exp}$\,=\,I(${E \parallel c}$)\,-\,I(${E \perp c}$) (green). (b) Simulated linear polarized XAS spectra of pure $|\pm j_z \rangle$ states. (c) Simulated linear dichroism of pure $|\pm j_z \rangle$ states, \textit{LD}$_{\rm jz}$, compared to experimental \textit{LD}$_{\rm exp}$ (green dotted curve taken from (a)).}
		\label{LD}
	\end{figure}
	
	\begin{figure*}[t]
		\begin{center}
			\includegraphics[width=1.1\columnwidth]{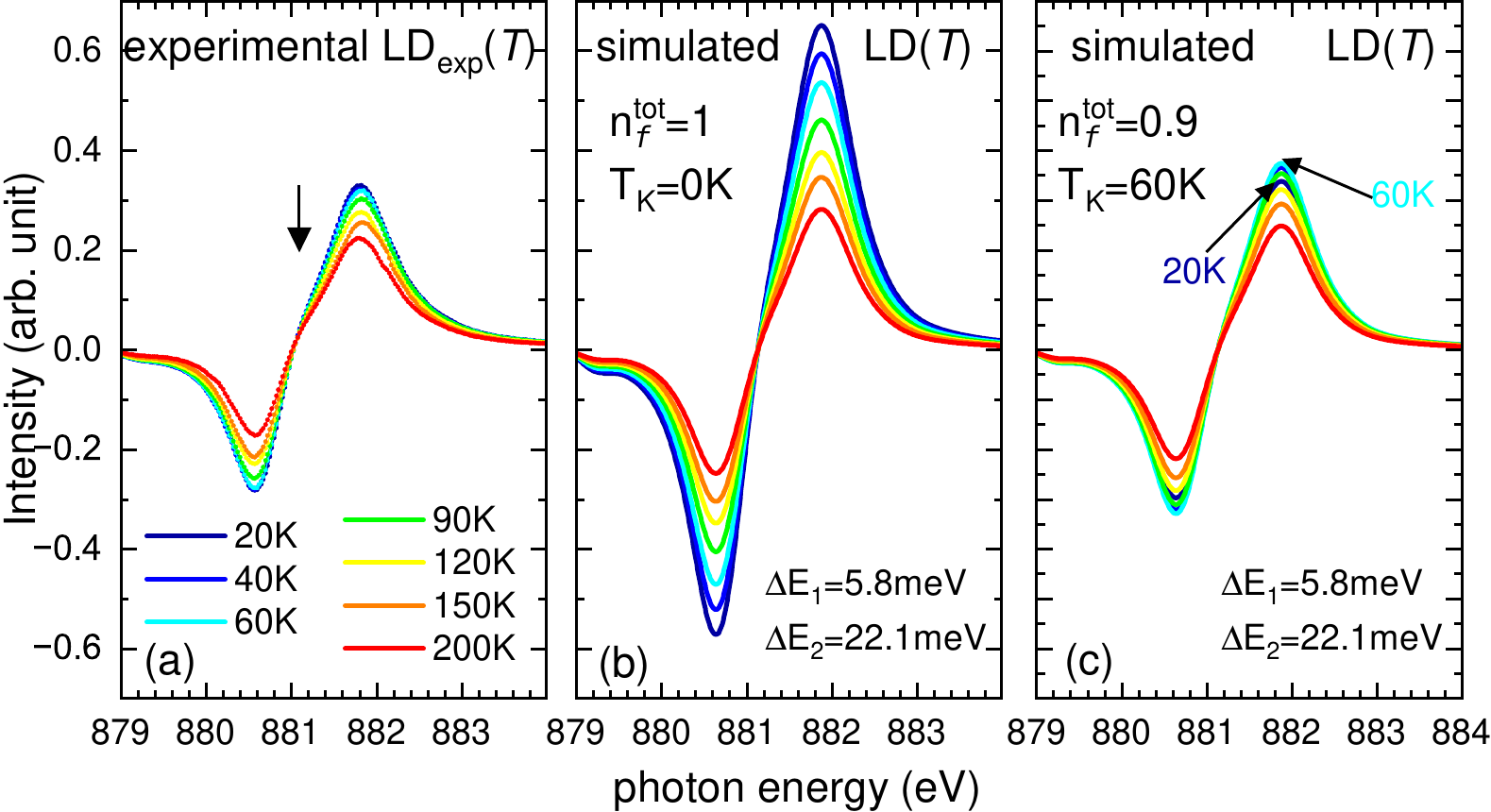}
			\includegraphics[width=0.88\columnwidth]{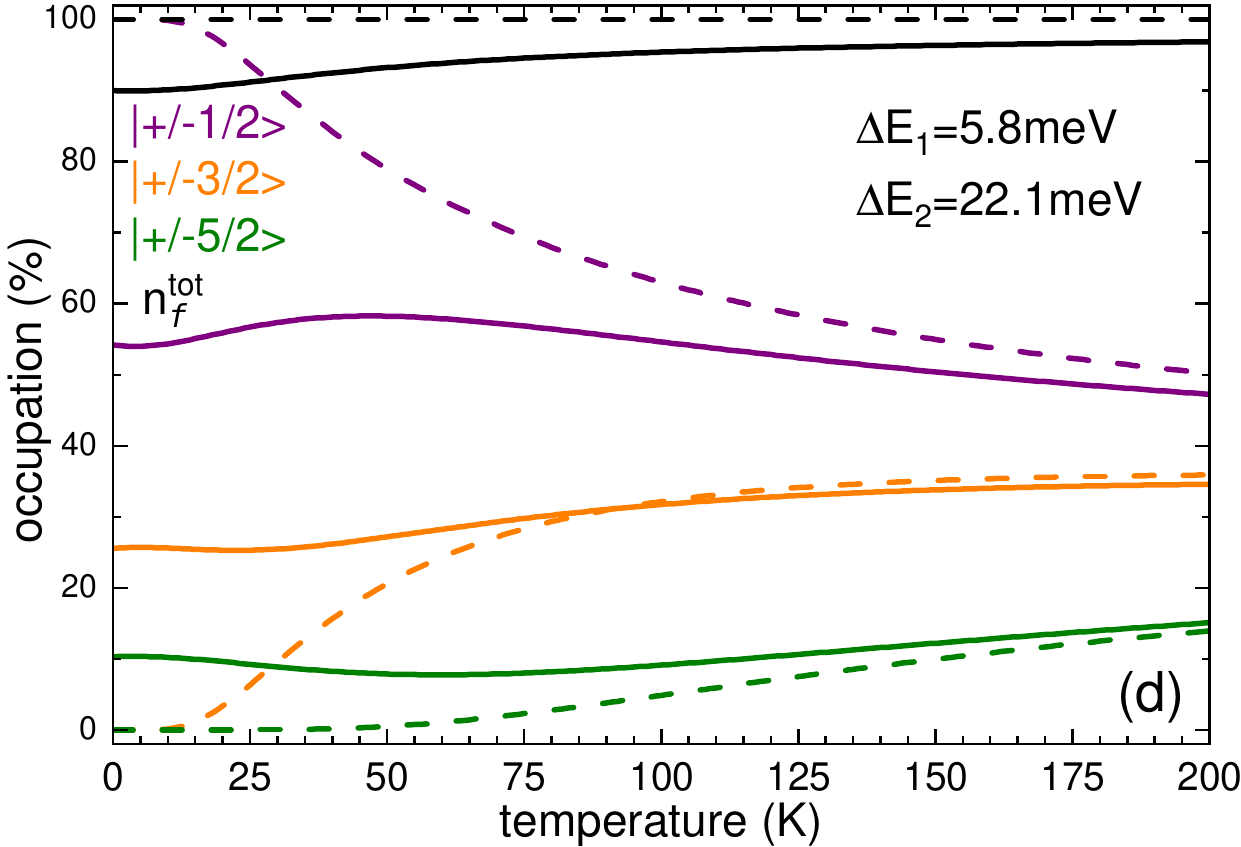}
		\end{center}
		\caption{(a) Temperature dependence of the experimental linear dichroism, \textit{LD}$_{\rm exp}$(\textit{T}). (b) Simulated \textit{LD}(\textit{T}) based on thermal occupation of states using the CEF model as described in the text, corresponding to T$_{\rm K}$\,=\,0\,K and \textit{n}$_f^{\rm tot}$(\textit{T})\,=\,1. (c) Simulated \textit{LD}(\textit{T}) taking into account the Kondo effect in addition to thermal occupation of states, see text. (d) Occupation of states as function of temperature for pure thermal occupation (dashed lines) and in the presence of the Kondo effect with T$_{\rm K}$\,=\,60\,K and \textit{n}$_f^{\rm tot}$(\textit{T}=0)\,=\,0.9 (solid lines). The black lines are the total occupation \textit{n}$_f^{\rm tot}$(\textit{T}) of the 4\textit{f} shell with (solid) and without  (dashed) Kondo effect. In the \textit{T}\,=\,0\,K limit, 35\% of the 4\textit{f} shell occupancy is due to $|\pm 3/2\rangle$ and $|\pm 5/2\rangle$ states.}
		\label{Tdep}
	\end{figure*}
	
   
	Figure\,\ref{Tdep}\,(a) shows the temperature dependence of the experimental linear dichroism, \textit{LD}$_{\rm exp}$(\textit{T}). The dichroism decreases continuously between 20 to 200K, from \textit{LD}$_{\rm exp}$(20\,K)\,=\,0.33 to \textit{LD}$_{\rm exp}$(200\,K)\,=\,0.22 (arbitrary units in our normalization), as measured at the energy of the maximum \textit{LD}$_{\rm exp}$(\textit{T}) at the \textit{M}$_5$ edge. This continuous decrease of \textit{LD}$_{\rm exp}$(\textit{T}) with increasing temperature confirms the sequence of CEF states, with the $|\pm 1/2\rangle$ lowest in energy and the $|\pm 5/2\rangle$ highest. By contrast,  if a $|\pm 5/2\rangle$ were close in energy to the $|\pm 1/2\rangle$ ground state, the \textit{LD} would change sign upon increasing  temperature. Likewise, if the $|\pm 3/2\rangle$ state were lowest in energy, the \textit{LD} would initially increase as the $|\pm 1/2\rangle$ becomes thermally populated, before decreasing as the $|\pm 5/2\rangle$ state becomes increasingly populated. Hence, the temperature dependence of \textit{\textit{LD}}$_{\rm exp}$(\textit{T}) independently confirms the sequence of CEF states deduced from the analysis of the static susceptibility\,\cite{Shu2021}.
	
	The temperature dependence of the dichroism was then calculated by quantitatively accounting only for thermal occupation of CEF states using the CEF scheme proposed in Ref.\,\cite{Shu2021}. The calculated dichroism is shown in  Fig.\,\ref{Tdep}\,(b). The dashed lines in Fig.\,\ref{Tdep}\,(d) indicate the corresponding thermal populations of the CEF levels. As expected from the qualitative comparison above, the calculated dichroism at 20\,K is significantly overestimated. It amounts to 0.66 at 20\,K and 0.28 at 200\,K in the same normalization. Thus, while the calculated value is in reasonable agreement with experiment at 200\,K, it exceeds the experimental dichroism at low temperatures by approximately a factor of two.

	 
	 The lack of reproducibility of \textit{\textit{LD}}$_{\rm exp}$ is reminiscent of the case observed in another ferromagnetic Kondo system, hexagonal CeRh$_3$B$_2$. In CeRh$_3B$$_2$, the $|\pm1/2 \rangle$ is also the lowest-energy CEF level, giving rise to an in-plane ordered magnetic moment that is strongly reduced\,\cite{Dhar1981,Galatanu2003}. In this compound, the experimentally observed \textit{LD}$_{\rm exp}$ is likewise smaller than the \textit{LD} calculated for a pure $|\pm1/2 \rangle$ state\,\cite{Amorese2023}. Furthermore, CeRh$_3$B$_2$ exhibits, like CeRh$_6$Ge$_4$, a seemingly contradictory set of observations indicative of a large Kondo temperature of several hundred Kelvin \,\cite{Malik1983,Maple1985,Kitaoka1985,Sampa1985,Shaheen1985,Fujimori1990,Allen1990,Takeuchi2004,Sundermann2016} despite its high magnetic ordering temperature\,\cite{Dhar1981,Galatanu2003}. This behavior has been explained in terms of a multiorbital ground state, consisting of the $|\pm1/2 \rangle$ component responsible for the magnetic order, and admixed higher lying CEF states that hybridize strongly with the conduction electrons and give rise to the Kondo effect\,\cite{Amorese2023}.

     In the following, a similar multiorbital scenario is investigated for CeRh$_6$Ge$_4$ using a simplified Anderson impurity model (SIAM) to calculate the occupation of the CEF states for a finite Kondo temperature and a total 4\textit{f} shell occupation \textit{n}$_f^{\rm tot}$\,$<$\,1. SIAM, treated within the non-crossing approximation (NCA)\,\cite{Zevin1988,Zwicknagl1990,Amorese2020}, accounts for the hybridization of the three CEF states with the conduction-electrons bath. Such a multiorbital ground state can generally occur when the Kondo energy scale and CEF splittings are comparable as in, e.g., CeRh$_3$B$_2$\cite{Amorese2023}, CeRh$_4$Sn$_6$\,\cite{Sundermann2015,Wissgott2016} or CeRh$_2$As$_2$\cite{Christovam2024}. The present calculations are based on the assumption of isotropic hybridization, whereby all CEF states are assumed to couple equally to the conduction-electron bath. In reality, however, symmetry considerations suggest that CEF states of different symmetry generally hybridize with different strengths.
     
     The SIAM/NCA calculations, assuming the same CEF model as above, were performed for \textit{n}$_f^{\rm tot}$\,=\,0.9 at \textit{T}\,=\,0\,K, as this value corresponds well to \textit{n}$_f^{\rm tot}$\,$\approx$\,0.95 at 200\,K as deduced from the analysis of the core-level data (see solid black line in Fig.\,\ref{Tdep}\,(d)). Calculations taking Kondo temperatures ranging from 0 to 190\,K found that \textit{T}$_{\rm K}$\,=\,60\,K provided the best overall agreement with experimentally determined LD. The Kondo effect induces a finite occupation of the excited CEF states even at \textit{T}\,=\,0\,K as demonstrated in Fig.\,\ref{Tdep}\,(d) by the solid lines. As a consequence, the partial occupation of the $|\pm 5/2\rangle$, which exhibits a large dichroism of opposite sign (see Fig.\,\ref{LD}\,(c)), strongly reduces the simulated \textit{LD} at low temperatures. The resulting Kondo-mixed ground state reproduces \textit{LD}$_{\rm exp}$ at 20\,K reasonably well (see Fig.\,\ref{Tdep}\,(c)). Furthermore, the experimental dichroism at 200\,K is well reproduced, as at this temperature the influence of the Kondo effect is strongly reduced and the occupations of the CEF states closely resemble those expected from purely thermal population in the absence of the Kondo effect. 
     
     However, a minor shortcoming is the initial increase of the simulated \textit{LD}. (For example, compare the relative maxima  at 20 K and 60 K in experimental data, Fig. 4(a), and the simulation, Fig. 4(c).)  This small discrepancy arises from the initially decreasing occupation of the excited states with increasing temperature due to diminishing impact of the Kondo effect as temperature rises (see Fig.\,\ref{Tdep}\,(d)). For even higher temperatures, thermal occupation becomes important, leading to a decrease of \textit{LD}. If the Kondo temperature of the excited $|\pm 5/2\rangle$ state were higher than that of the $|\pm 1/2\rangle$, the influence of the Kondo effect would persist to higher temperatures and this non-monotonic behavior of the \textit{LD} could potentially be avoided. However, different Kondo hybridization strengths for the individual CEF states are not included in the present model.
     
     A multiorbital ground state with contributions from $|\pm 3/2\rangle$ and $|\pm 5/2\rangle$, each having a different, and possibly stronger hybridization than the $|\pm 1/2\rangle$ component, accounts for the discrepancy between a Kondo temperatures deduced from specific heat \textit{C} (\textit{T}$_{\rm K}^C$\,$\sim$\,20\,K) under the assumption of a spin-1/2 groundstate\,\cite{Matsuoka2015} and a broad spin response found in inelastic neutron scattering\,\cite{Shu2021}. For example, the INS data cannot be described using the crystal-field model derived from static susceptibility together with an intrinsic line broadening corresponding to the Kondo temperature obtained from specific heat. Instead, the INS data are consistent with a Kondo temperature of $\sim$60\,K. This is demonstrated in the Appendix, where calculations of the imaginary part of the dynamical susceptibility and corresponding structure factors of CeRh$_6$Ge$_4$ are presented within the framework of the present Anderson impurity model, thereby explicitly taking into account the multiorbital nature of the ground state.
     
     A multiorbital ground state also is implied from a Kadowaki-Woods ratio close to that expected for a 4-fold degenerate ground state\,\cite{Shen2020}. To give the measured Sommerfeld coefficient, in this limit \textit{T}$_{\rm K}$ should be about three times larger than expected for an effective S\,=\,1/2 state, i.e. T$_{\rm K}$\,$\sim$\,60\,K\,\cite{Rajan1983}. This same value of \textit{T}$_{\rm K}$\,$\sim$\,60\,K, determined independently in our SIAM/NCA calculations of LD, also  accounts for the magnitude and general temperature dependence of LD$_{\rm exp}$(T). Given approximations in both determinations of \textit{T}$_{\rm K}$, it is not possible to say that the ground state is 4-fold degenerate, but it clearly is effectively greater than doubly degenerate due to multiorbital contributions.   
        
     In SIAM/NCA calculations, the Kondo hybridization was taken to be orbital independent, but there is already evidence of a Kondo effect at 200\,K in the core-level PES as well as XAS data (Fig.\,\ref{PES} and Fig.\,\ref{ISO}), which is well above \textit{T}$_{\rm K}$\,$\sim$\,60\,K resulting from an orbital averaged hybridization and which is comparable to the energy of the highest lying CEF doublet. Though ARPES was performed only to 120\,K, these measurements also revealed clear evidence of \textit{f-c} hybridization and its anisotropy at 120\,K\,\cite{Wu2021}. Thus, both XAS and (AR)PES argue for a Kondo effect on the $|\pm 5/2\rangle$ CEF level with a characteristic temperature greater than the orbital-average \textit{T}$_{\rm K}$. As mentioned, a stronger hybridization of the $|\pm 5/2\rangle$ level would improve agreement between experiment and the calculated temperature dependence of the LD.
     
     Though the $|\pm 1/2\rangle$ orbital dictates in-plane ordered moments, Kondo-derived  mixing of higher lying CEF levels into the ground state plays a role in reducing the magnitude of the ordered moment and the \textit{f}-shell occupation. It is not surprising, then, that angle-dependent deHaas-vanAlphen measurements give a Fermi surface topology that has both localized and itinerant 4\textit{f} characters found in LDA\,+\,U band calculations, but the Fermi volume is more characteristic of the 4\textit{f}-localized limit\,\cite{Wang2021}, which supports a Kondo-breakdown interpretation of quantum criticality in CeRh$_6$Ge$_4$\,\cite{Shen2020}. Duality in 4\textit{f} character arising from multiorbital mixing does not rule out this view of criticality; however, it leaves open alternative interpretations of its origin\,\cite{Zhan2025,Thomas2024}.
     
     Pressure dependence of crystal fields may play a role in an alternative interpretation\,\cite{Thomas2024}. Results presented here argue that this suggestion might merit more attention, particularly because Kondo and CEF energy scales in CeRh$_6$Ge$_4$ are comparable. Typically, the Kondo scale in Ce-based Kondo systems is strongly volume dependent and gives a dominant,  low-temperature contribution to a large Gruneissen parameter. The CEF splitting also is volume dependent but much less so in the absence of a Kondo effect in the higher lying CEF states\,\cite{Thompson1994}. As we have discussed, charge and magnetic spectroscopies, however, do show the presence of a Kondo effect in these CEF states, and therefore the measured Gruneissen parameter is a mixture of both contributions with the CEF-derived Gruneissen parameter making a more significant contribution, even at temperatures well below the first excited CEF level. Such a CEF contribution could influence an interpretation of the diverging temperature dependence of the Gruneissen parameter of CeRh$_6$Ge$_4$ as it is tuned to a QCP\,\cite{Zhan2025}.
       
     \section{Summary}
	 In summary, core-level PES and isotropic XAS measurements provide clear evidence of a temperature-dependent 4\textit{f}$^0$ contribution to the electronic spectrum of CeRh$_6$Ge$_4$ that persists to at least 200\,K, even at atmospheric pressure.  Linearly polarized XAS reveals a temperature-dependent linear dichroism consistent with the sequence of CEF states inferred from the static susceptibility, $|\pm1/2 \rangle$, $|\pm3/2 \rangle$ and $|\pm5/2 \rangle$. The experimental dichroism cannot be reproduced by an ionic full-multiplet model alone, but is captured when the Kondo effect is included, assuming a total 4\textit{f}-shell occupation of 0.9 and an orbital-independent Kondo temperature of 60\,K resulting from orbital independent hybridization. This value of \textit{T}$_{\rm K}$ also is consistent with a quasi-quartet ground state inferred from thermodynamic and transport measurements.	 The detailed temperature dependence of the linear dichroism, however, deviates from that expected for isotropic hybridization, suggesting stronger hybridization for the excited CEF states. This suggestion aligns with conclusions from ARPES and neutron scattering experiments. Importantly, the Kondo effect induces non-trivial contributions of both first and second excited crystal-field doublets into the ground state. The resulting multiorbital ground state should be considered the starting point for a model of pressure-induced criticality. Our experiments do not rule out a Kondo-breakdown interpretation of quantum criticality in CeRh$_6$Ge$_4$, but they do leave open alternative scenarios and in particular highlight the possible role of crystal fields in an interpretation.

\section{Appendix}
\subsection{Isotropic and directional dependend XAS spectra}

\begin{figure}[t]
	\begin{center}
		\includegraphics[width=0.97\columnwidth]{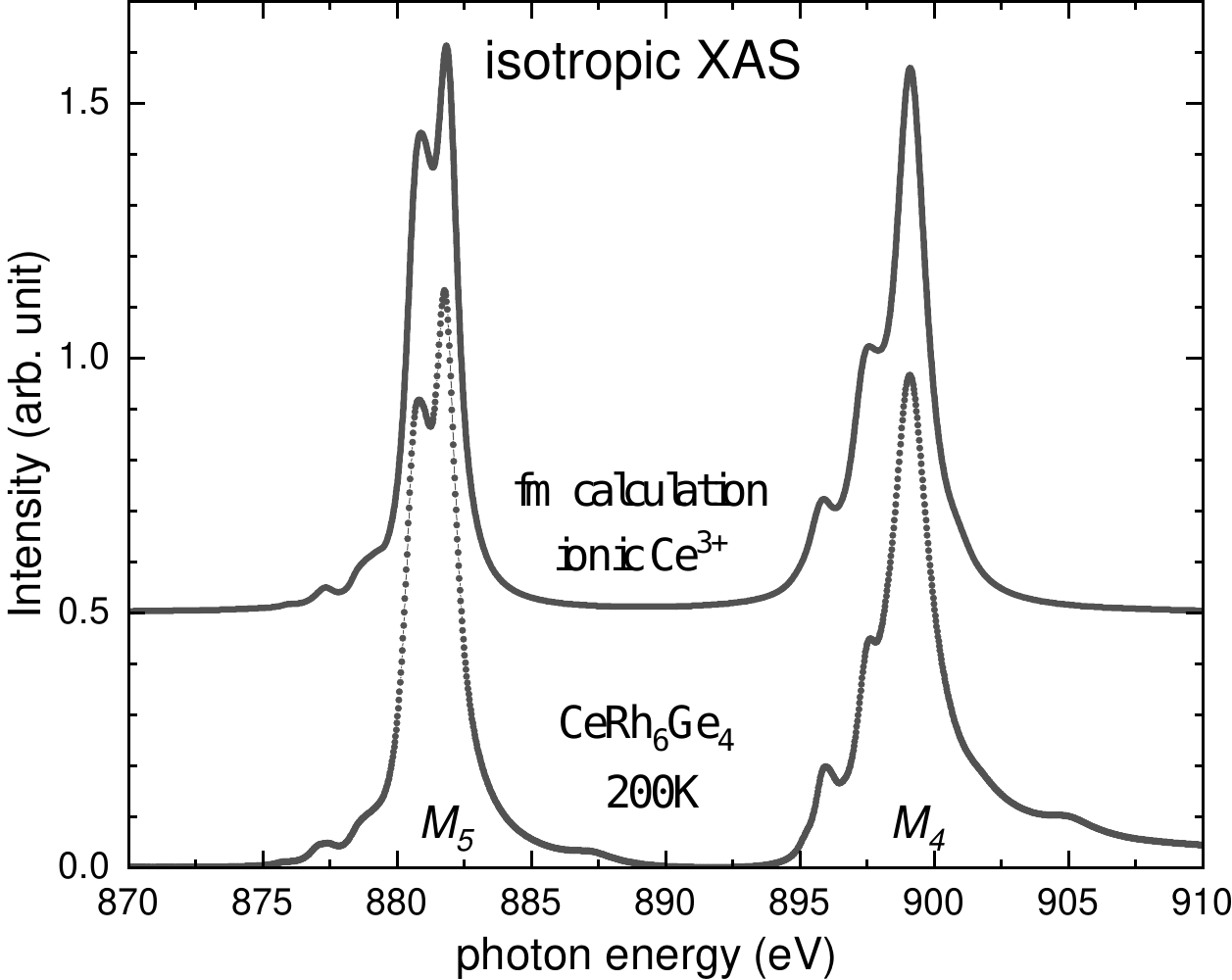}
	\end{center}
	\caption{Isotropic XAS spectra at the Ce \textit{M}$_{4,5}$ edges of CeRh$_6$Ge$_4$ at 200\,K constructed from linear polarized data based on Eq.\,\ref{eq_corr_c} and full-multiplet (\textit{fm}) calculation for Ce$^{3+}$ (top).  }
	\label{A1}
\end{figure}

\begin{figure}[t]
	\begin{center}
		\includegraphics[width=0.97\columnwidth]{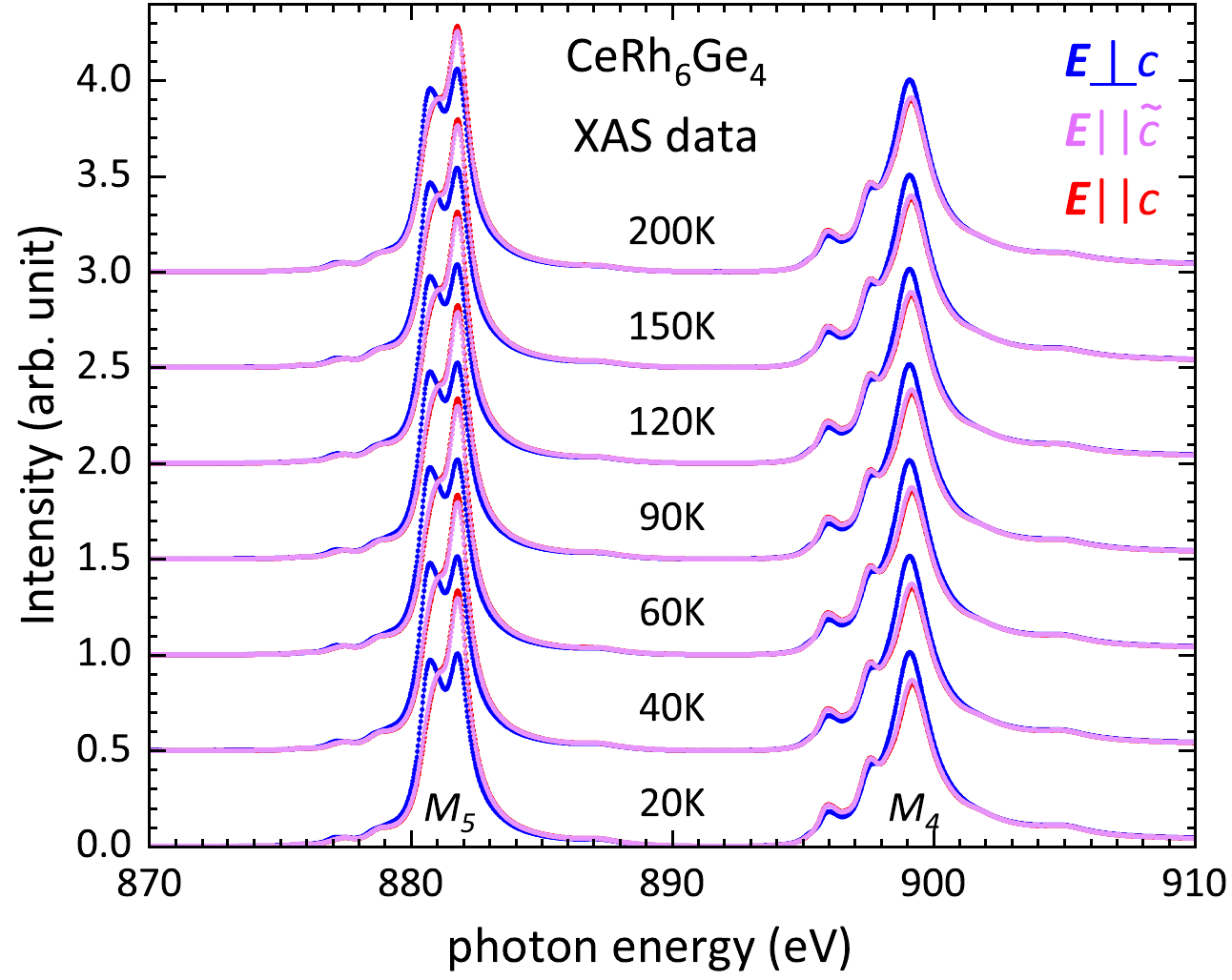}
	\end{center}
	\caption{Linearly polarized XAS spectra of CeRh$_6$Ge$_4$ at various temperature for the electric field vector $\vec{E}$\,$\perp$\,\textit{c} (blue) and $\vec{E}$\,$\parallel$\,$\tilde{\textit{c}}$ (pink) with $\tilde{\textit{c}}$ 20$^{\circ}$ off the crystallographic \textit{c} direction. The spectra $\vec{E}$\,$\parallel$\,$\textit{c}$ (red) were reconstructed following eq.\,(1).}
	\label{A2}
\end{figure}

Figure\,\ref{A1} shows the agreement between our isotropic simulation and XAS data. The small satellites at 887\,eV and 905\,eV are not reproduced, as they arise from the Ce$^{4+}$ configuration in the ground state and correspond to the transition (3\textit{d}$^{10}$4\textit{f}$^0$\,$\rightarrow$\,3\textit{d}$^{9}$4\textit{f}$^1$) which is not accounted for in the ionic calculation.

Due to geometrical constraints in the experiment, measurements with $\vec{E}$\,$\parallel$\,$c$ could not be realized, Instead, the spectra were recorded with $\vec{E}$\,$\parallel$\,$\tilde{\textit{c}}$, where $\tilde{\textit{c}}$ is tilted by 20$^{\circ}$ with respect to the crystallographic \textit{c} direction. Within the dipole limit of XAS in $D_{\rm 3h}$ point symmetry, the $\vec{E}$\,$\parallel$\,$c$ spectrum can nevertheless be reconstructed as a linear combination of $\vec{E}$\,$\parallel$\,$\tilde{\textit{c}}$ and $\vec{E}$\,$\perp$\,$c$ according to the relation

\begin{equation}\label{eq_corr_c}
I(\vec{E}\parallel\tilde{\textit{c}}) = \text{cos}^2(20^{\circ}) \cdot I(\vec{E}\parallel{c}) + \text{sin}^2(20^{\circ}) \cdot I(\vec{E}\perp{c}).
\end{equation}

Figure\,\ref{A2} shows the measured XAS spectra for $\vec{E}$\,$\perp$\,$c$ and $\vec{E}$\,$\parallel$\,$\tilde{\textit{c}}$, together with the reconstructed $\vec{E}$\,$\parallel$\,$c$, at several temperatures between 20 and 200\,K. The latter differs noticeably only in the energy regions where the \textit{LD} is large. 
Accordingly, the isotropic spectrum shown in Fig.\,\ref{A1} is given by I$_{\rm iso}$\,$\approx$\,0.38$\cdot$I($\vec{E}$\,$\parallel$\,$\tilde{\textit{c}}$)\,+\,0.62$\cdot$I($\vec{E}$\,$\perp$\,$c$), and the \textit{LD}, I($\vec{E}$\,$\parallel$\,$c$)\,-\,I($\vec{E}$\,$\perp$\,$c$), is approximately 13\% larger than the difference between the measured spectra, I($\vec{E}$\,$\parallel$\,$\tilde{\textit{c}}$)\,-\,I($\vec{E}$\,$\perp$\,$c$).

\section{Appendix}
\subsection{Imaginary part of dynamical susceptibility and structure factor}
The INS data of CeRh$_6$Ge$_4$ are not compatible with the line broadening expected for the specific-heat-derived Kondo temperature of 20\,K nor with a Kondo temperature well above 100\,K as the width of the high energy tail in the neutron data may suggest\,\cite{Shu2021}. 

In the following, the dynamical susceptibility Im\,$\chi\left(\omega;T\right)$ and the corresponding structure factor are calculated  for the following parameters
\begin{itemize}
	\item Kondo temperature $T_{K}=60K$
	\item Low-temperature 4\textit{f}-valence $n_{f}\left(T=0\right)=0.9$
	\item CEF scheme:\medskip{}
	\begin{table}[H]
		\begin{centering}
			\begin{tabular}{c|c|c}
				\hline 
				state & energy (meV) & energy (K)\tabularnewline
				\hline 
				\hline 
				$\left|\pm\frac{1}{2}\right\rangle $ & 0 & 0\tabularnewline
				\hline 
				$\left|\pm\frac{3}{2}\right\rangle $ & 5.8 & 67.3\tabularnewline
				\hline 
				$\left|\pm\frac{5}{2}\right\rangle $ & 22.1 & 256.5\tabularnewline
				\hline 
			\end{tabular}
			\par\end{centering}
		\caption{CEF states and excitation energies\label{tab:CEFStatesEnergies}}
\end{table}
\item Hybridization strength:\\
The \textit{isotropic} hybridization strength is adjusted so as to reproduce the low-temperature 4\textit{f}-valence.
\end{itemize}

\begin{figure}[t]
	\begin{center}
		\includegraphics[width=0.55\columnwidth]{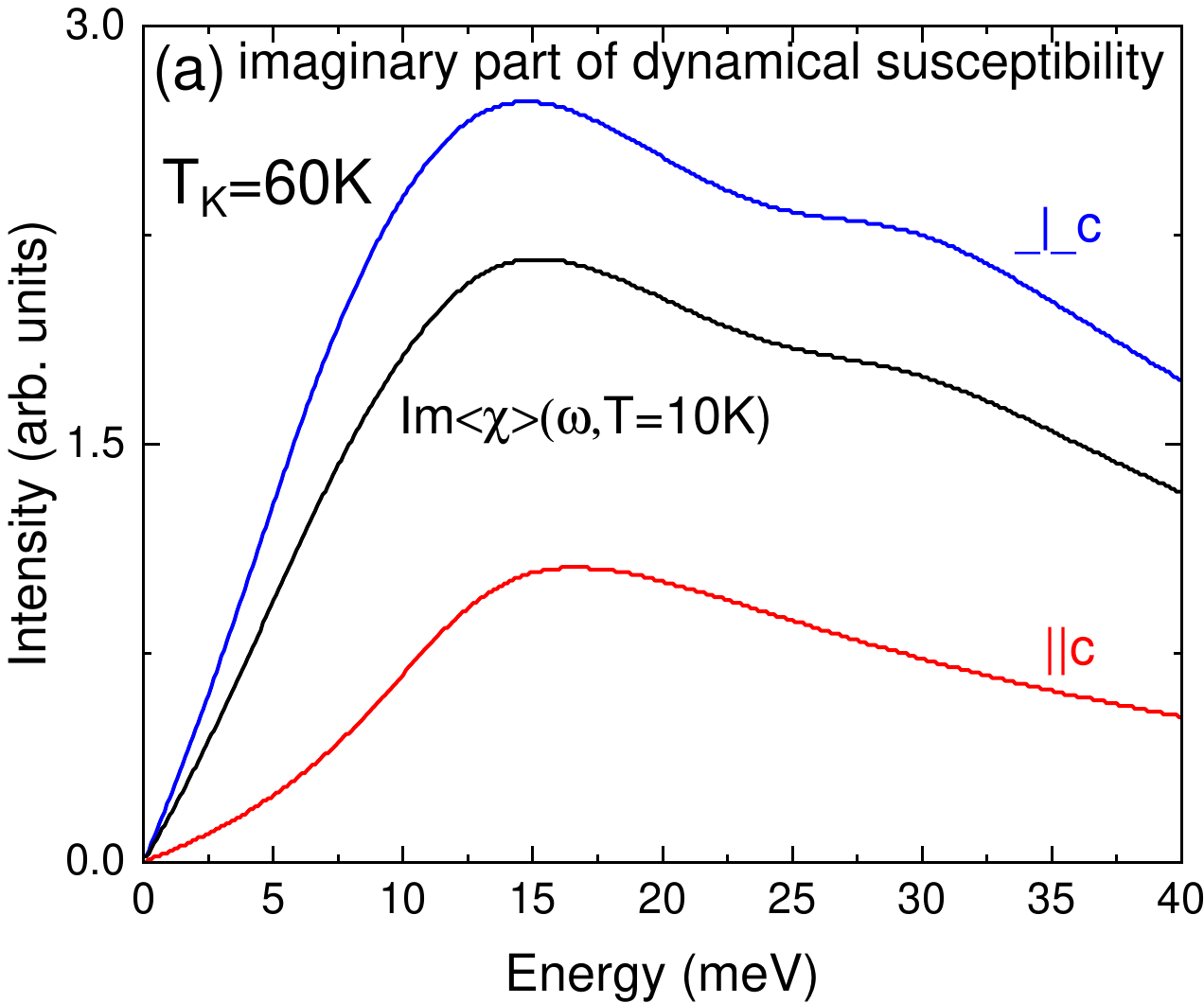}
		\includegraphics[width=0.55\columnwidth]{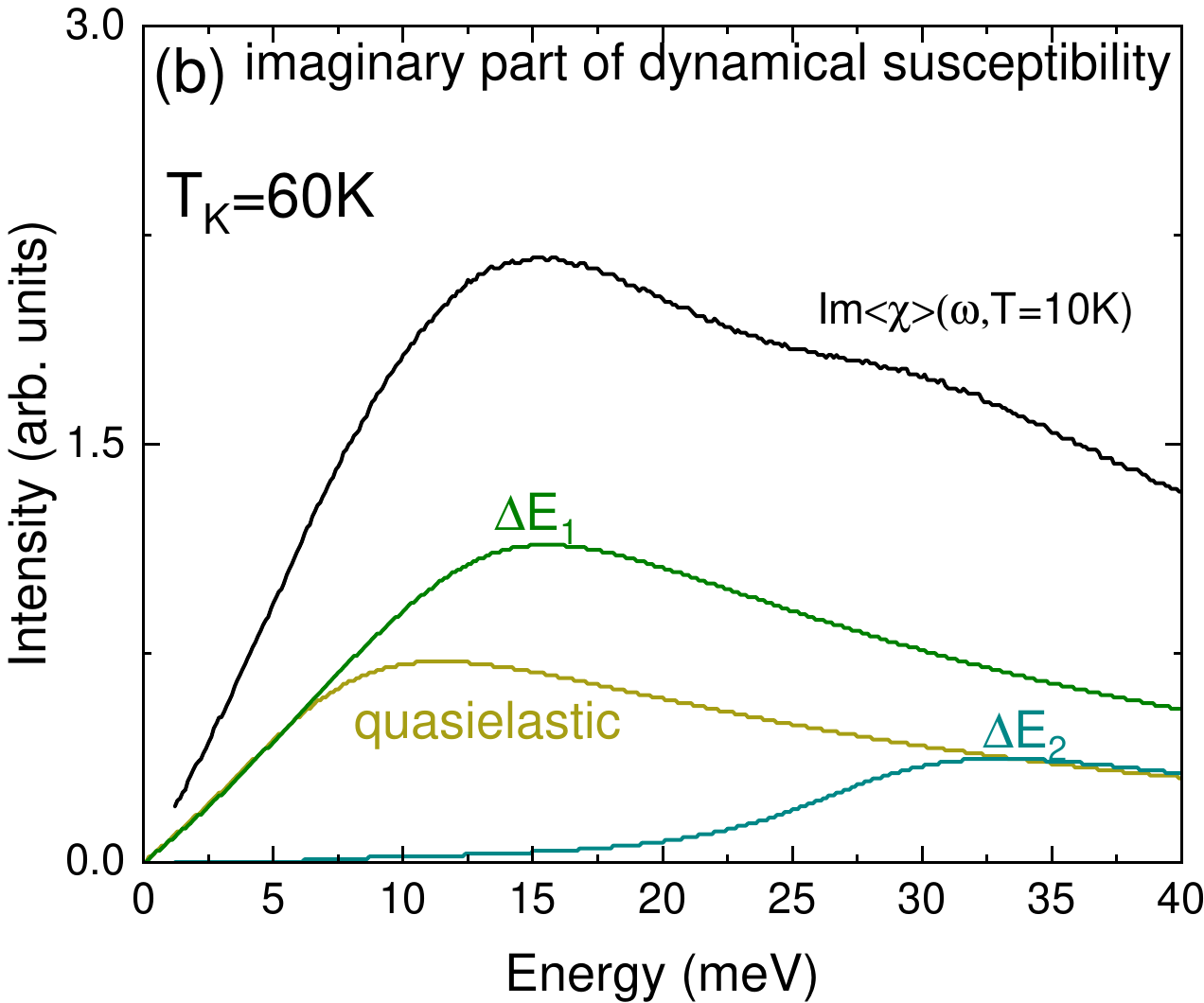}
		\includegraphics[width=0.55\columnwidth]{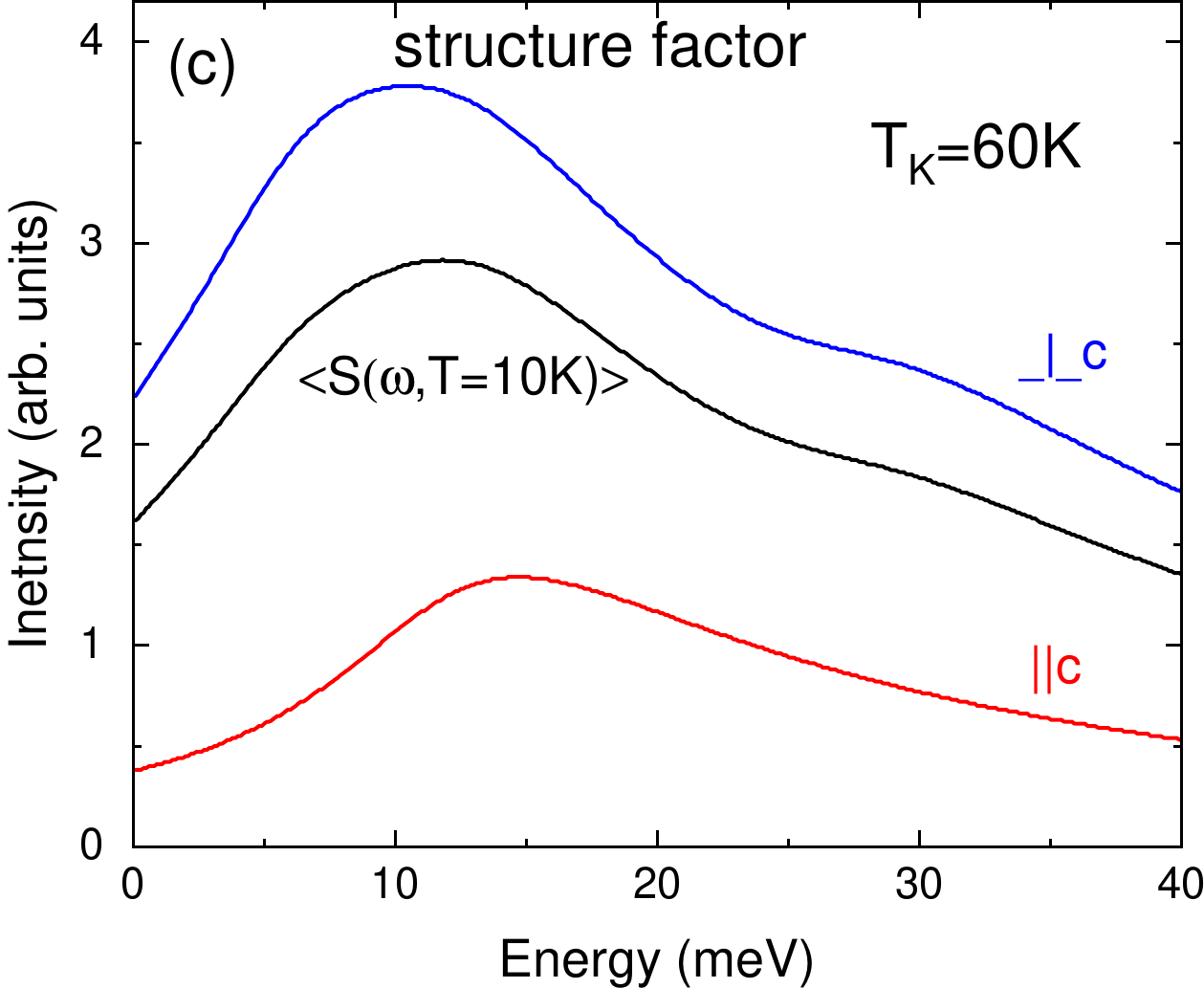}
	\end{center}
	\caption{Calculated spin response of CeRh$_6$Ge$_4$: Directional dependence (a) and spectral contributions (b) of imaginary part of the dynamical susceptibility Im\,$\left\langle \chi\left(\omega,T\right)\right\rangle $ and corresponding structure factor $\left\langle S\left(\omega;T\right)\right\rangle $ (c).}
	\label{Chi_S}
\end{figure}

As a result of Kondo-singlet formation, \textit{f}-contributions from the excited CEF states are mixed into the ground state. For isotropic
hybridization, the relative weights of these contributions depend upon the ratio of the Kondo temperature and the CEF excitation energies. In the resulting multiorbital ground state, with contributions of $\left|\pm\frac{1}{2}\right\rangle $, $\left|\pm\frac{3}{2}\right\rangle $ and $\left|\pm\frac{5}{2}\right\rangle $ as in Fig.\,\ref{Tdep}\,(d), the dipole selection rules for INS allow transitions not only from the ground state to the first excited state at 5.8\,meV, but also to the second excited state at 22.1\,meV. Furthermore, the excitation energies appear shifted by the Kondo temperature, i.e., $\Delta$E$_{\rm CEF}$\,+\,T$_{\rm K}$.

We show the dynamical susceptibilities Im\,$\chi_{\perp c}\left(\omega,T\right)$,
Im\,$\chi_{\parallel c}\left(\omega,T\right)$, and the orientationally averaged Im\,$\left\langle \chi\left(\omega,T\right)\right\rangle $\,=\,Im\,$\left(\frac{2}{3}\chi_{\perp c}\left(\omega;T\right)+\frac{1}{3}\chi_{\parallel c}\left(\omega;T\right)\right)$ in Fig.\,\ref{Chi_S}\,(a) as well as the respective contributions from the Kondo ground state (quasielastic) and van Vleck terms ($\Delta$E$_1$ and $\Delta$E$_2$) in Fig.\,\ref{Chi_S}\,(b), all for isotorpic hybridization. The corresponding structure factors $\left\langle S\left(\omega;T\right)\right\rangle $ are shown in Fig\,\ref{Chi_S}\,(c). In the Kondo regime, the calculated structure factor reproduces the characteristic features of the INS data quite well\,\cite{Shu2021}, namely a rather broad peak around 10\,meV, a slight bump between 20-30\,meV, and an extended high energy tail. The latter is a consequence of Kondo-induced broadening.

\section{Acknowledgment}
We thank Z. Riedel for sharing results of his specific heat measurements on CeRh$_6$Ge$_4$ and LaRh$_6$Ge$_4$. A.S. acknowledges support from the German Research Foundation (DFG) - grant N$^{\circ}$ 567326535. Work at Los Alamos National Laboratory was performed under the auspices of the U.S. Department of Energy, Office of Basic Energy Sciences, Division of Materials Science and Engineering under project “Quantum Fluctuations in Narrow-Band Systems”. All authors acknowledge the support from the Max Planck-POSTECH-Hsinchu Center for Complex Phase Materials.

%


\begin{thebibliography}{50}%
	\makeatletter
	\providecommand \@ifxundefined [1]{%
		\@ifx{#1\undefined}
	}%
	\providecommand \@ifnum [1]{%
		\ifnum #1\expandafter \@firstoftwo
		\else \expandafter \@secondoftwo
		\fi
	}%
	\providecommand \@ifx [1]{%
		\ifx #1\expandafter \@firstoftwo
		\else \expandafter \@secondoftwo
		\fi
	}%
	\providecommand \natexlab [1]{#1}%
	\providecommand \enquote  [1]{``#1''}%
	\providecommand \bibnamefont  [1]{#1}%
	\providecommand \bibfnamefont [1]{#1}%
	\providecommand \citenamefont [1]{#1}%
	\providecommand \href@noop [0]{\@secondoftwo}%
	\providecommand \href [0]{\begingroup \@sanitize@url \@href}%
	\providecommand \@href[1]{\@@startlink{#1}\@@href}%
	\providecommand \@@href[1]{\endgroup#1\@@endlink}%
	\providecommand \@sanitize@url [0]{\catcode `\\12\catcode `\$12\catcode
		`\&12\catcode `\#12\catcode `\^12\catcode `\_12\catcode `\%12\relax}%
	\providecommand \@@startlink[1]{}%
	\providecommand \@@endlink[0]{}%
	\providecommand \url  [0]{\begingroup\@sanitize@url \@url }%
	\providecommand \@url [1]{\endgroup\@href {#1}{\urlprefix }}%
	\providecommand \urlprefix  [0]{URL }%
	\providecommand \Eprint [0]{\href }%
	\providecommand \doibase [0]{https://doi.org/}%
	\providecommand \selectlanguage [0]{\@gobble}%
	\providecommand \bibinfo  [0]{\@secondoftwo}%
	\providecommand \bibfield  [0]{\@secondoftwo}%
	\providecommand \translation [1]{[#1]}%
	\providecommand \BibitemOpen [0]{}%
	\providecommand \bibitemStop [0]{}%
	\providecommand \bibitemNoStop [0]{.\EOS\space}%
	\providecommand \EOS [0]{\spacefactor3000\relax}%
	\providecommand \BibitemShut  [1]{\csname bibitem#1\endcsname}%
	\let\auto@bib@innerbib\@empty
	\bibitem [{\citenamefont {Belitz}\ \emph {et~al.}(1999)\citenamefont {Belitz},
		\citenamefont {Kirkpatrick},\ and\ \citenamefont {Vojta}}]{Belitz1999}%
	\BibitemOpen
	\bibfield  {author} {\bibinfo {author} {\bibfnamefont {D.}~\bibnamefont
			{Belitz}}, \bibinfo {author} {\bibfnamefont {T.~R.}\ \bibnamefont
			{Kirkpatrick}},\ and\ \bibinfo {author} {\bibfnamefont {T.}~\bibnamefont
			{Vojta}},\ }\bibfield  {title} {\bibinfo {title} {First order transitions and
			multicritical points in weak itinerant ferromagnets},\ }\href
	{https://doi.org/10.1103/PhysRevLett.82.4707} {\bibfield  {journal} {\bibinfo
			{journal} {Phys. Rev. Lett.}\ }\textbf {\bibinfo {volume} {82}},\ \bibinfo
		{pages} {4707} (\bibinfo {year} {1999})}\BibitemShut {NoStop}%
	\bibitem [{\citenamefont {Chubukov}\ \emph {et~al.}(2004)\citenamefont
		{Chubukov}, \citenamefont {P\'epin},\ and\ \citenamefont
		{Rech}}]{Chubukov2004}%
	\BibitemOpen
	\bibfield  {author} {\bibinfo {author} {\bibfnamefont {A.~V.}\ \bibnamefont
			{Chubukov}}, \bibinfo {author} {\bibfnamefont {C.}~\bibnamefont {P\'epin}},\
		and\ \bibinfo {author} {\bibfnamefont {J.}~\bibnamefont {Rech}},\ }\bibfield
	{title} {\bibinfo {title} {Instability of the quantum-critical point of
			itinerant ferromagnets},\ }\href
	{https://doi.org/10.1103/PhysRevLett.92.147003} {\bibfield  {journal}
		{\bibinfo  {journal} {Phys. Rev. Lett.}\ }\textbf {\bibinfo {volume} {92}},\
		\bibinfo {pages} {147003} (\bibinfo {year} {2004})}\BibitemShut {NoStop}%
	\bibitem [{\citenamefont {Belitz}\ \emph {et~al.}(2005)\citenamefont {Belitz},
		\citenamefont {Kirkpatrick},\ and\ \citenamefont {Vojta}}]{Belitz2005}%
	\BibitemOpen
	\bibfield  {author} {\bibinfo {author} {\bibfnamefont {D.}~\bibnamefont
			{Belitz}}, \bibinfo {author} {\bibfnamefont {T.~R.}\ \bibnamefont
			{Kirkpatrick}},\ and\ \bibinfo {author} {\bibfnamefont {T.}~\bibnamefont
			{Vojta}},\ }\bibfield  {title} {\bibinfo {title} {How generic scale
			invariance influences quantum and classical phase transitions},\ }\href
	{https://doi.org/10.1103/RevModPhys.77.579} {\bibfield  {journal} {\bibinfo
			{journal} {Rev. Mod. Phys.}\ }\textbf {\bibinfo {volume} {77}},\ \bibinfo
		{pages} {579} (\bibinfo {year} {2005})}\BibitemShut {NoStop}%
	\bibitem [{\citenamefont {Conduit}\ \emph {et~al.}(2009)\citenamefont
		{Conduit}, \citenamefont {Green},\ and\ \citenamefont
		{Simons}}]{Conduit2009}%
	\BibitemOpen
	\bibfield  {author} {\bibinfo {author} {\bibfnamefont {G.~J.}\ \bibnamefont
			{Conduit}}, \bibinfo {author} {\bibfnamefont {A.~G.}\ \bibnamefont {Green}},\
		and\ \bibinfo {author} {\bibfnamefont {B.~D.}\ \bibnamefont {Simons}},\
	}\bibfield  {title} {\bibinfo {title} {Inhomogeneous phase formation on the
			border of itinerant ferromagnetism},\ }\href
	{https://doi.org/10.1103/PhysRevLett.103.207201} {\bibfield  {journal}
		{\bibinfo  {journal} {Phys. Rev. Lett.}\ }\textbf {\bibinfo {volume} {103}},\
		\bibinfo {pages} {207201} (\bibinfo {year} {2009})}\BibitemShut {NoStop}%
	\bibitem [{\citenamefont {Brando}\ \emph {et~al.}(2016)\citenamefont {Brando},
		\citenamefont {Belitz}, \citenamefont {Grosche},\ and\ \citenamefont
		{Kirkpatrick}}]{Brando2016}%
	\BibitemOpen
	\bibfield  {author} {\bibinfo {author} {\bibfnamefont {M.}~\bibnamefont
			{Brando}}, \bibinfo {author} {\bibfnamefont {D.}~\bibnamefont {Belitz}},
		\bibinfo {author} {\bibfnamefont {F.~M.}\ \bibnamefont {Grosche}},\ and\
		\bibinfo {author} {\bibfnamefont {T.~R.}\ \bibnamefont {Kirkpatrick}},\
	}\bibfield  {title} {\bibinfo {title} {Metallic quantum ferromagnets},\
	}\href {https://doi.org/10.1103/RevModPhys.88.025006} {\bibfield  {journal}
		{\bibinfo  {journal} {Rev. Mod. Phys.}\ }\textbf {\bibinfo {volume} {88}},\
		\bibinfo {pages} {025006} (\bibinfo {year} {2016})}\BibitemShut {NoStop}%
	\bibitem [{\citenamefont {Kirkpatrick}\ and\ \citenamefont
		{Belitz}(2020)}]{Kirkpatrick2020}%
	\BibitemOpen
	\bibfield  {author} {\bibinfo {author} {\bibfnamefont {T.~R.}\ \bibnamefont
			{Kirkpatrick}}\ and\ \bibinfo {author} {\bibfnamefont {D.}~\bibnamefont
			{Belitz}},\ }\bibfield  {title} {\bibinfo {title} {Ferromagnetic quantum
			critical point in noncentrosymmetric systems},\ }\href
	{https://doi.org/10.1103/PhysRevLett.124.147201} {\bibfield  {journal}
		{\bibinfo  {journal} {Phys. Rev. Lett.}\ }\textbf {\bibinfo {volume} {124}},\
		\bibinfo {pages} {147201} (\bibinfo {year} {2020})}\BibitemShut {NoStop}%
	\bibitem [{\citenamefont {Steppke}\ \emph {et~al.}(2013)\citenamefont
		{Steppke}, \citenamefont {Küchler}, \citenamefont {Lausberg}, \citenamefont
		{Lengyel}, \citenamefont {Steinke}, \citenamefont {Borth}, \citenamefont
		{Lühmann}, \citenamefont {Krellner}, \citenamefont {Nicklas}, \citenamefont
		{Geibel}, \citenamefont {Steglich},\ and\ \citenamefont
		{Brando}}]{Steppke2013}%
	\BibitemOpen
	\bibfield  {author} {\bibinfo {author} {\bibfnamefont {A.}~\bibnamefont
			{Steppke}}, \bibinfo {author} {\bibfnamefont {R.}~\bibnamefont {Küchler}},
		\bibinfo {author} {\bibfnamefont {S.}~\bibnamefont {Lausberg}}, \bibinfo
		{author} {\bibfnamefont {E.}~\bibnamefont {Lengyel}}, \bibinfo {author}
		{\bibfnamefont {L.}~\bibnamefont {Steinke}}, \bibinfo {author} {\bibfnamefont
			{R.}~\bibnamefont {Borth}}, \bibinfo {author} {\bibfnamefont
			{T.}~\bibnamefont {Lühmann}}, \bibinfo {author} {\bibfnamefont
			{C.}~\bibnamefont {Krellner}}, \bibinfo {author} {\bibfnamefont
			{M.}~\bibnamefont {Nicklas}}, \bibinfo {author} {\bibfnamefont
			{C.}~\bibnamefont {Geibel}}, \bibinfo {author} {\bibfnamefont
			{F.}~\bibnamefont {Steglich}},\ and\ \bibinfo {author} {\bibfnamefont
			{M.}~\bibnamefont {Brando}},\ }\bibfield  {title} {\bibinfo {title}
		{Ferromagnetic quantum critical point in the heavy-fermion metal
			{YbNi$_4$(P$_{1-x}$As$_x$)$_2$}},\ }\href
	{https://doi.org/10.1126/science.1230583} {\bibfield  {journal} {\bibinfo
			{journal} {Science}\ }\textbf {\bibinfo {volume} {339}},\ \bibinfo {pages}
		{933} (\bibinfo {year} {2013})},\ \Eprint
	{https://arxiv.org/abs/https://www.science.org/doi/pdf/10.1126/science.1230583}
	{https://www.science.org/doi/pdf/10.1126/science.1230583} \BibitemShut
	{NoStop}%
	\bibitem [{\citenamefont {Vo{\ss}winkel}\ \emph {et~al.}(2012)\citenamefont
		{Vo{\ss}winkel}, \citenamefont {Niehaus}, \citenamefont {Rodewald},\ and\
		\citenamefont {P{\"o}ttgen}}]{Vosswinkel2012}%
	\BibitemOpen
	\bibfield  {author} {\bibinfo {author} {\bibfnamefont {D.}~\bibnamefont
			{Vo{\ss}winkel}}, \bibinfo {author} {\bibfnamefont {O.}~\bibnamefont
			{Niehaus}}, \bibinfo {author} {\bibfnamefont {U.~C.}\ \bibnamefont
			{Rodewald}},\ and\ \bibinfo {author} {\bibfnamefont {R.}~\bibnamefont
			{P{\"o}ttgen}},\ }\bibfield  {title} {\bibinfo {title} {Bismuth flux growth
			of {CeRh$_6$Ge$_4$} and {CeRh$_2$Ge$_2$} single crystals},\ }\href
	{https://doi.org/doi:10.5560/znb.2012-0265} {\bibfield  {journal} {\bibinfo
			{journal} {Zeitschrift für Naturforschung B}\ }\textbf {\bibinfo {volume}
			{67}},\ \bibinfo {pages} {1241} (\bibinfo {year} {2012})}\BibitemShut
	{NoStop}%
	\bibitem [{\citenamefont {Matsuoka}\ \emph {et~al.}(2015)\citenamefont
		{Matsuoka}, \citenamefont {Hondo}, \citenamefont {Fujii}, \citenamefont
		{Oshima}, \citenamefont {Sugawara}, \citenamefont {Sakurai}, \citenamefont
		{Ohta}, \citenamefont {Kneidinger}, \citenamefont {Salamakha}, \citenamefont
		{Michor},\ and\ \citenamefont {Bauer}}]{Matsuoka2015}%
	\BibitemOpen
	\bibfield  {author} {\bibinfo {author} {\bibfnamefont {E.}~\bibnamefont
			{Matsuoka}}, \bibinfo {author} {\bibfnamefont {C.}~\bibnamefont {Hondo}},
		\bibinfo {author} {\bibfnamefont {T.}~\bibnamefont {Fujii}}, \bibinfo
		{author} {\bibfnamefont {A.}~\bibnamefont {Oshima}}, \bibinfo {author}
		{\bibfnamefont {H.}~\bibnamefont {Sugawara}}, \bibinfo {author}
		{\bibfnamefont {T.}~\bibnamefont {Sakurai}}, \bibinfo {author} {\bibfnamefont
			{H.}~\bibnamefont {Ohta}}, \bibinfo {author} {\bibfnamefont {F.}~\bibnamefont
			{Kneidinger}}, \bibinfo {author} {\bibfnamefont {L.}~\bibnamefont
			{Salamakha}}, \bibinfo {author} {\bibfnamefont {H.}~\bibnamefont {Michor}},\
		and\ \bibinfo {author} {\bibfnamefont {E.}~\bibnamefont {Bauer}},\ }\bibfield
	{title} {\bibinfo {title} {Ferromagnetic transition at 2.5 {K} in the
			hexagonal {K}ondo-lattice compound {CeRh$_6$Ge$_4$}},\ }\href
	{https://doi.org/10.7566/JPSJ.84.073704} {\bibfield  {journal} {\bibinfo
			{journal} {J. Phys. Soc. Jpn.}\ }\textbf {\bibinfo {volume} {84}},\ \bibinfo
		{pages} {073704} (\bibinfo {year} {2015})},\ \Eprint
	{https://arxiv.org/abs/https://doi.org/10.7566/JPSJ.84.073704}
	{https://doi.org/10.7566/JPSJ.84.073704} \BibitemShut {NoStop}%
	\bibitem [{\citenamefont {Kotegawa}\ \emph {et~al.}(2019)\citenamefont
		{Kotegawa}, \citenamefont {Matsuoka}, \citenamefont {Uga}, \citenamefont
		{Takemura}, \citenamefont {Manago}, \citenamefont {Chikuchi}, \citenamefont
		{Sugawara}, \citenamefont {Tou},\ and\ \citenamefont
		{Harima}}]{Kotegawa2019}%
	\BibitemOpen
	\bibfield  {author} {\bibinfo {author} {\bibfnamefont {H.}~\bibnamefont
			{Kotegawa}}, \bibinfo {author} {\bibfnamefont {E.}~\bibnamefont {Matsuoka}},
		\bibinfo {author} {\bibfnamefont {T.}~\bibnamefont {Uga}}, \bibinfo {author}
		{\bibfnamefont {M.}~\bibnamefont {Takemura}}, \bibinfo {author}
		{\bibfnamefont {M.}~\bibnamefont {Manago}}, \bibinfo {author} {\bibfnamefont
			{N.}~\bibnamefont {Chikuchi}}, \bibinfo {author} {\bibfnamefont
			{H.}~\bibnamefont {Sugawara}}, \bibinfo {author} {\bibfnamefont
			{H.}~\bibnamefont {Tou}},\ and\ \bibinfo {author} {\bibfnamefont
			{H.}~\bibnamefont {Harima}},\ }\bibfield  {title} {\bibinfo {title}
		{Indication of ferromagnetic quantum critical point in {K}ondo lattice
			{CeRh$_6$Ge$_4$}},\ }\href {https://doi.org/10.7566/JPSJ.88.093702}
	{\bibfield  {journal} {\bibinfo  {journal} {J. Phys. Soc. Jpn.}\ }\textbf
		{\bibinfo {volume} {88}},\ \bibinfo {pages} {093702} (\bibinfo {year}
		{2019})},\ \Eprint
	{https://arxiv.org/abs/https://doi.org/10.7566/JPSJ.88.093702}
	{https://doi.org/10.7566/JPSJ.88.093702} \BibitemShut {NoStop}%
	\bibitem [{\citenamefont {Shen}\ \emph {et~al.}(2020)\citenamefont {Shen},
		\citenamefont {Zhang}, \citenamefont {Komijani}, \citenamefont {Nicklas},
		\citenamefont {Borth}, \citenamefont {Wang}, \citenamefont {Chen},
		\citenamefont {Nie}, \citenamefont {Li}, \citenamefont {Lu}, \citenamefont
		{Lee}, \citenamefont {Smidman}, \citenamefont {Steglich}, \citenamefont
		{Coleman},\ and\ \citenamefont {Yuan}}]{Shen2020}%
	\BibitemOpen
	\bibfield  {author} {\bibinfo {author} {\bibfnamefont {B.}~\bibnamefont
			{Shen}}, \bibinfo {author} {\bibfnamefont {Y.}~\bibnamefont {Zhang}},
		\bibinfo {author} {\bibfnamefont {Y.}~\bibnamefont {Komijani}}, \bibinfo
		{author} {\bibfnamefont {M.}~\bibnamefont {Nicklas}}, \bibinfo {author}
		{\bibfnamefont {R.}~\bibnamefont {Borth}}, \bibinfo {author} {\bibfnamefont
			{A.}~\bibnamefont {Wang}}, \bibinfo {author} {\bibfnamefont {Y.}~\bibnamefont
			{Chen}}, \bibinfo {author} {\bibfnamefont {Z.}~\bibnamefont {Nie}}, \bibinfo
		{author} {\bibfnamefont {R.}~\bibnamefont {Li}}, \bibinfo {author}
		{\bibfnamefont {X.}~\bibnamefont {Lu}}, \bibinfo {author} {\bibfnamefont
			{H.}~\bibnamefont {Lee}}, \bibinfo {author} {\bibfnamefont {M.}~\bibnamefont
			{Smidman}}, \bibinfo {author} {\bibfnamefont {F.}~\bibnamefont {Steglich}},
		\bibinfo {author} {\bibfnamefont {P.}~\bibnamefont {Coleman}},\ and\ \bibinfo
		{author} {\bibfnamefont {H.}~\bibnamefont {Yuan}},\ }\bibfield  {title}
	{\bibinfo {title} {Strange-metal behaviour in a pure ferromagnetic {K}ondo
			lattice},\ }\href {https://doi.org/10.1038/s41586-020-2052-z} {\bibfield
		{journal} {\bibinfo  {journal} {Nature}\ }\textbf {\bibinfo {volume} {579}},\
		\bibinfo {pages} {51} (\bibinfo {year} {2020})}\BibitemShut {NoStop}%
	\bibitem [{\citenamefont {Paschen}\ and\ \citenamefont
		{Si}(2020)}]{Paschen2020}%
	\BibitemOpen
	\bibfield  {author} {\bibinfo {author} {\bibfnamefont {S.}~\bibnamefont
			{Paschen}}\ and\ \bibinfo {author} {\bibfnamefont {Q.}~\bibnamefont {Si}},\
	}\bibfield  {title} {\bibinfo {title} {Quantum phases driven by strong
			correlations},\ }\href {https://doi.org/10.1038/s42254-020-00262-6}
	{\bibfield  {journal} {\bibinfo  {journal} {Nat. Rev. Phys.}\ }\textbf
		{\bibinfo {volume} {3}},\ \bibinfo {pages} {9} (\bibinfo {year}
		{2020})}\BibitemShut {NoStop}%
	\bibitem [{\citenamefont {Komijani}\ and\ \citenamefont
		{Coleman}(2018)}]{Komijani2018}%
	\BibitemOpen
	\bibfield  {author} {\bibinfo {author} {\bibfnamefont {Y.}~\bibnamefont
			{Komijani}}\ and\ \bibinfo {author} {\bibfnamefont {P.}~\bibnamefont
			{Coleman}},\ }\bibfield  {title} {\bibinfo {title} {Model for a ferromagnetic
			quantum critical point in a 1d {K}ondo lattice},\ }\href
	{https://doi.org/10.1103/PhysRevLett.120.157206} {\bibfield  {journal}
		{\bibinfo  {journal} {Phys. Rev. Lett.}\ }\textbf {\bibinfo {volume} {120}},\
		\bibinfo {pages} {157206} (\bibinfo {year} {2018})}\BibitemShut {NoStop}%
	\bibitem [{\citenamefont {Shin}\ \emph {et~al.}(2024)\citenamefont {Shin},
		\citenamefont {Ramires}, \citenamefont {Pomjakushin}, \citenamefont
		{Plokhikh},\ and\ \citenamefont {Pomjakushina}}]{Shin2024}%
	\BibitemOpen
	\bibfield  {author} {\bibinfo {author} {\bibfnamefont {S.}~\bibnamefont
			{Shin}}, \bibinfo {author} {\bibfnamefont {A.}~\bibnamefont {Ramires}},
		\bibinfo {author} {\bibfnamefont {V.}~\bibnamefont {Pomjakushin}}, \bibinfo
		{author} {\bibfnamefont {I.}~\bibnamefont {Plokhikh}},\ and\ \bibinfo
		{author} {\bibfnamefont {E.}~\bibnamefont {Pomjakushina}},\ }\bibfield
	{title} {\bibinfo {title} {Ferromagnetic quantum critical point protected by
			nonsymmorphic symmetry in a {K}ondo metal},\ }\href
	{https://doi.org/10.1038/s41467-024-52720-9} {\bibfield  {journal} {\bibinfo
			{journal} {Nat. Commun.}\ }\textbf {\bibinfo {volume} {15}},\ \bibinfo
		{pages} {8423} (\bibinfo {year} {2024})}\BibitemShut {NoStop}%
	\bibitem [{\citenamefont {Yamamoto}\ and\ \citenamefont
		{Si}(2010)}]{Yamamoto2010}%
	\BibitemOpen
	\bibfield  {author} {\bibinfo {author} {\bibfnamefont {S.~J.}\ \bibnamefont
			{Yamamoto}}\ and\ \bibinfo {author} {\bibfnamefont {Q.}~\bibnamefont {Si}},\
	}\bibfield  {title} {\bibinfo {title} {Metallic ferromagnetism in the {K}ondo
			lattice},\ }\href {https://doi.org/10.1073/pnas.1009498107} {\bibfield
		{journal} {\bibinfo  {journal} {Proc. Nat. Acad. Science}\ }\textbf {\bibinfo
			{volume} {107}},\ \bibinfo {pages} {15704} (\bibinfo {year} {2010})},\
	\Eprint
	{https://arxiv.org/abs/https://www.pnas.org/doi/pdf/10.1073/pnas.1009498107}
	{https://www.pnas.org/doi/pdf/10.1073/pnas.1009498107} \BibitemShut {NoStop}%
	\bibitem [{\citenamefont {Zhan}\ \emph {et~al.}(2025)\citenamefont {Zhan},
		\citenamefont {Zhang}, \citenamefont {Zhang}, \citenamefont {Liu},
		\citenamefont {Nie}, \citenamefont {Chen}, \citenamefont {Jiao},
		\citenamefont {Komijani}, \citenamefont {Smidman}, \citenamefont {Steglich},
		\citenamefont {Coleman},\ and\ \citenamefont {Yuan}}]{Zhan2025}%
	\BibitemOpen
	\bibfield  {author} {\bibinfo {author} {\bibfnamefont {J.}~\bibnamefont
			{Zhan}}, \bibinfo {author} {\bibfnamefont {Y.}~\bibnamefont {Zhang}},
		\bibinfo {author} {\bibfnamefont {J.}~\bibnamefont {Zhang}}, \bibinfo
		{author} {\bibfnamefont {Y.}~\bibnamefont {Liu}}, \bibinfo {author}
		{\bibfnamefont {Z.}~\bibnamefont {Nie}}, \bibinfo {author} {\bibfnamefont
			{Y.}~\bibnamefont {Chen}}, \bibinfo {author} {\bibfnamefont {L.}~\bibnamefont
			{Jiao}}, \bibinfo {author} {\bibfnamefont {Y.}~\bibnamefont {Komijani}},
		\bibinfo {author} {\bibfnamefont {M.}~\bibnamefont {Smidman}}, \bibinfo
		{author} {\bibfnamefont {F.}~\bibnamefont {Steglich}}, \bibinfo {author}
		{\bibfnamefont {P.}~\bibnamefont {Coleman}},\ and\ \bibinfo {author}
		{\bibfnamefont {H.}~\bibnamefont {Yuan}},\ }\bibfield  {title} {\bibinfo
		{title} {Critical fluctuations and conserved dynamics in a strange
			ferromagnetic metal},\ }\href {https://doi.org/10.1103/bp6l-46z7} {\bibfield
		{journal} {\bibinfo  {journal} {Phys. Rev. Lett.}\ }\textbf {\bibinfo
			{volume} {135}},\ \bibinfo {pages} {266504} (\bibinfo {year}
		{2025})}\BibitemShut {NoStop}%
	\bibitem [{\citenamefont {Wang}\ \emph {et~al.}(2021)\citenamefont {Wang},
		\citenamefont {Du}, \citenamefont {Zhang}, \citenamefont {Graf},
		\citenamefont {Shen}, \citenamefont {Chen}, \citenamefont {Liu},
		\citenamefont {Smidman}, \citenamefont {Cao}, \citenamefont {Steglich},\ and\
		\citenamefont {Yuan}}]{Wang2021}%
	\BibitemOpen
	\bibfield  {author} {\bibinfo {author} {\bibfnamefont {A.}~\bibnamefont
			{Wang}}, \bibinfo {author} {\bibfnamefont {F.}~\bibnamefont {Du}}, \bibinfo
		{author} {\bibfnamefont {Y.}~\bibnamefont {Zhang}}, \bibinfo {author}
		{\bibfnamefont {D.}~\bibnamefont {Graf}}, \bibinfo {author} {\bibfnamefont
			{B.}~\bibnamefont {Shen}}, \bibinfo {author} {\bibfnamefont {Y.}~\bibnamefont
			{Chen}}, \bibinfo {author} {\bibfnamefont {Y.}~\bibnamefont {Liu}}, \bibinfo
		{author} {\bibfnamefont {M.}~\bibnamefont {Smidman}}, \bibinfo {author}
		{\bibfnamefont {C.}~\bibnamefont {Cao}}, \bibinfo {author} {\bibfnamefont
			{F.}~\bibnamefont {Steglich}},\ and\ \bibinfo {author} {\bibfnamefont
			{H.}~\bibnamefont {Yuan}},\ }\bibfield  {title} {\bibinfo {title} {Localized
			4f-electrons in the quantum critical heavy fermion ferromagnet
			{CeRh$_6$Ge$_4$}},\ }\href
	{https://doi.org/https://doi.org/10.1016/j.scib.2021.03.006} {\bibfield
		{journal} {\bibinfo  {journal} {Science Bulletin}\ }\textbf {\bibinfo
			{volume} {66}},\ \bibinfo {pages} {1389} (\bibinfo {year}
		{2021})}\BibitemShut {NoStop}%
	\bibitem [{\citenamefont {Wu}\ \emph {et~al.}(2021)\citenamefont {Wu},
		\citenamefont {Zhang}, \citenamefont {Du}, \citenamefont {Shen},
		\citenamefont {Zheng}, \citenamefont {Fang}, \citenamefont {Smidman},
		\citenamefont {Cao}, \citenamefont {Steglich}, \citenamefont {Yuan},
		\citenamefont {Denlinger},\ and\ \citenamefont {Liu}}]{Wu2021}%
	\BibitemOpen
	\bibfield  {author} {\bibinfo {author} {\bibfnamefont {Y.}~\bibnamefont
			{Wu}}, \bibinfo {author} {\bibfnamefont {Y.}~\bibnamefont {Zhang}}, \bibinfo
		{author} {\bibfnamefont {F.}~\bibnamefont {Du}}, \bibinfo {author}
		{\bibfnamefont {B.}~\bibnamefont {Shen}}, \bibinfo {author} {\bibfnamefont
			{H.}~\bibnamefont {Zheng}}, \bibinfo {author} {\bibfnamefont
			{Y.}~\bibnamefont {Fang}}, \bibinfo {author} {\bibfnamefont {M.}~\bibnamefont
			{Smidman}}, \bibinfo {author} {\bibfnamefont {C.}~\bibnamefont {Cao}},
		\bibinfo {author} {\bibfnamefont {F.}~\bibnamefont {Steglich}}, \bibinfo
		{author} {\bibfnamefont {H.}~\bibnamefont {Yuan}}, \bibinfo {author}
		{\bibfnamefont {J.~D.}\ \bibnamefont {Denlinger}},\ and\ \bibinfo {author}
		{\bibfnamefont {Y.}~\bibnamefont {Liu}},\ }\bibfield  {title} {\bibinfo
		{title} {Anisotropic $c\ensuremath{-}f$ hybridization in the ferromagnetic
			quantum critical metal ${\mathrm{cerh}}_{6}{\mathrm{ge}}_{4}$},\ }\href
	{https://doi.org/10.1103/PhysRevLett.126.216406} {\bibfield  {journal}
		{\bibinfo  {journal} {Phys. Rev. Lett.}\ }\textbf {\bibinfo {volume} {126}},\
		\bibinfo {pages} {216406} (\bibinfo {year} {2021})}\BibitemShut {NoStop}%
	\bibitem [{\citenamefont {Thomas}\ \emph {et~al.}(2024)\citenamefont {Thomas},
		\citenamefont {Seo}, \citenamefont {Asaba}, \citenamefont {Ronning},
		\citenamefont {Rosa}, \citenamefont {Bauer},\ and\ \citenamefont
		{Thompson}}]{Thomas2024}%
	\BibitemOpen
	\bibfield  {author} {\bibinfo {author} {\bibfnamefont {S.~M.}\ \bibnamefont
			{Thomas}}, \bibinfo {author} {\bibfnamefont {S.}~\bibnamefont {Seo}},
		\bibinfo {author} {\bibfnamefont {T.}~\bibnamefont {Asaba}}, \bibinfo
		{author} {\bibfnamefont {F.}~\bibnamefont {Ronning}}, \bibinfo {author}
		{\bibfnamefont {P.~F.~S.}\ \bibnamefont {Rosa}}, \bibinfo {author}
		{\bibfnamefont {E.~D.}\ \bibnamefont {Bauer}},\ and\ \bibinfo {author}
		{\bibfnamefont {J.~D.}\ \bibnamefont {Thompson}},\ }\bibfield  {title}
	{\bibinfo {title} {Probing quantum criticality in ferromagnetic
			{CeRh}$_6${Ge}$_4$},\ }\href {https://doi.org/10.1103/PhysRevB.109.L121105}
	{\bibfield  {journal} {\bibinfo  {journal} {Phys. Rev. B}\ }\textbf {\bibinfo
			{volume} {109}},\ \bibinfo {pages} {L121105} (\bibinfo {year}
		{2024})}\BibitemShut {NoStop}%
	\bibitem [{\citenamefont {Itokazu}\ \emph {et~al.}(2025)\citenamefont
		{Itokazu}, \citenamefont {Kirikoshi}, \citenamefont {Jeschke},\ and\
		\citenamefont {Otsuki}}]{Itokazu2025}%
	\BibitemOpen
	\bibfield  {author} {\bibinfo {author} {\bibfnamefont {S.}~\bibnamefont
			{Itokazu}}, \bibinfo {author} {\bibfnamefont {A.}~\bibnamefont {Kirikoshi}},
		\bibinfo {author} {\bibfnamefont {H.~O.}\ \bibnamefont {Jeschke}},\ and\
		\bibinfo {author} {\bibfnamefont {J.}~\bibnamefont {Otsuki}},\ }\bibfield
	{title} {\bibinfo {title} {From localized 4f electrons to anisotropic
			exchange interactions in ferromagnetic {CeRh$_6$Ge$_4$}},\ }\bibfield
	{journal} {\bibinfo  {journal} {Commun. Mater.}\ }\textbf {\bibinfo {volume}
		{6}},\ \href {https://doi.org/10.1038/s43246-025-00998-7}
	{10.1038/s43246-025-00998-7} (\bibinfo {year} {2025})\BibitemShut {NoStop}%
	\bibitem [{\citenamefont {Shu}\ \emph {et~al.}(2021)\citenamefont {Shu},
		\citenamefont {Adroja}, \citenamefont {Hillier}, \citenamefont {Zhang},
		\citenamefont {Chen}, \citenamefont {Shen}, \citenamefont {Orlandi},
		\citenamefont {Walker}, \citenamefont {Liu}, \citenamefont {Cao},
		\citenamefont {Steglich}, \citenamefont {Yuan},\ and\ \citenamefont
		{Smidman}}]{Shu2021}%
	\BibitemOpen
	\bibfield  {author} {\bibinfo {author} {\bibfnamefont {J.~W.}\ \bibnamefont
			{Shu}}, \bibinfo {author} {\bibfnamefont {D.~T.}\ \bibnamefont {Adroja}},
		\bibinfo {author} {\bibfnamefont {A.~D.}\ \bibnamefont {Hillier}}, \bibinfo
		{author} {\bibfnamefont {Y.~J.}\ \bibnamefont {Zhang}}, \bibinfo {author}
		{\bibfnamefont {Y.~X.}\ \bibnamefont {Chen}}, \bibinfo {author}
		{\bibfnamefont {B.}~\bibnamefont {Shen}}, \bibinfo {author} {\bibfnamefont
			{F.}~\bibnamefont {Orlandi}}, \bibinfo {author} {\bibfnamefont {H.~C.}\
			\bibnamefont {Walker}}, \bibinfo {author} {\bibfnamefont {Y.}~\bibnamefont
			{Liu}}, \bibinfo {author} {\bibfnamefont {C.}~\bibnamefont {Cao}}, \bibinfo
		{author} {\bibfnamefont {F.}~\bibnamefont {Steglich}}, \bibinfo {author}
		{\bibfnamefont {H.~Q.}\ \bibnamefont {Yuan}},\ and\ \bibinfo {author}
		{\bibfnamefont {M.}~\bibnamefont {Smidman}},\ }\bibfield  {title} {\bibinfo
		{title} {Magnetic order and crystalline electric field excitations of the
			quantum critical heavy-fermion ferromagnet {CeRh$_6$Ge$_4$}},\ }\href
	{https://doi.org/10.1103/PhysRevB.104.L140411} {\bibfield  {journal}
		{\bibinfo  {journal} {Phys. Rev. B}\ }\textbf {\bibinfo {volume} {104}},\
		\bibinfo {pages} {L140411} (\bibinfo {year} {2021})}\BibitemShut {NoStop}%
	\bibitem [{\citenamefont {Yamamoto}\ \emph {et~al.}(2025)\citenamefont
		{Yamamoto}, \citenamefont {Park}, \citenamefont {Riedel}, \citenamefont
		{Sherpa}, \citenamefont {Thompson}, \citenamefont {Ronning}, \citenamefont
		{Bauer}, \citenamefont {Dioguardi},\ and\ \citenamefont
		{Hirata}}]{Yamamoto2025}%
	\BibitemOpen
	\bibfield  {author} {\bibinfo {author} {\bibfnamefont {R.}~\bibnamefont
			{Yamamoto}}, \bibinfo {author} {\bibfnamefont {S.}~\bibnamefont {Park}},
		\bibinfo {author} {\bibfnamefont {Z.~W.}\ \bibnamefont {Riedel}}, \bibinfo
		{author} {\bibfnamefont {P.}~\bibnamefont {Sherpa}}, \bibinfo {author}
		{\bibfnamefont {J.~D.}\ \bibnamefont {Thompson}}, \bibinfo {author}
		{\bibfnamefont {F.}~\bibnamefont {Ronning}}, \bibinfo {author} {\bibfnamefont
			{E.~D.}\ \bibnamefont {Bauer}}, \bibinfo {author} {\bibfnamefont {A.~P.}\
			\bibnamefont {Dioguardi}},\ and\ \bibinfo {author} {\bibfnamefont
			{M.}~\bibnamefont {Hirata}},\ }\bibfield  {title} {\bibinfo {title} {Evidence
			for easy-plane xy ferromagnetism in heavy-fermion quantum-critical
			{CeRh}$_6${Ge}$_4$},\ }\href {https://doi.org/10.1103/z17k-jz1s} {\bibfield
		{journal} {\bibinfo  {journal} {Phys. Rev. B}\ }\textbf {\bibinfo {volume}
			{112}},\ \bibinfo {pages} {155152} (\bibinfo {year} {2025})}\BibitemShut
	{NoStop}%
	\bibitem [{\citenamefont {Haverkort}(2016)}]{Haverkort2016}%
	\BibitemOpen
	\bibfield  {author} {\bibinfo {author} {\bibfnamefont {M.~W.}\ \bibnamefont
			{Haverkort}},\ }\bibfield  {title} {\bibinfo {title} {\textit{Quanty} for
			core level spectroscopy - excitons, resonances and band excitations in time
			and frequency domain},\ }\href
	{https://doi.org/10.1088/1742-6596/712/1/012001} {\bibfield  {journal}
		{\bibinfo  {journal} {J. Phys. Conf. Ser.}\ }\textbf {\bibinfo {volume}
			{712}},\ \bibinfo {pages} {012001} (\bibinfo {year} {2016})}\BibitemShut
	{NoStop}%
	\bibitem [{\citenamefont {Imer}\ and\ \citenamefont
		{Wuilloud}(1987)}]{Imer1987}%
	\BibitemOpen
	\bibfield  {author} {\bibinfo {author} {\bibfnamefont {J.-M.}\ \bibnamefont
			{Imer}}\ and\ \bibinfo {author} {\bibfnamefont {E.}~\bibnamefont
			{Wuilloud}},\ }\bibfield  {title} {\bibinfo {title} {A simple model
			calculation for {XPS, BIS and EELS} 4f-excitations in ce and la compounds},\
	}\href {https://doi.org/10.1007/BF01311650} {\bibfield  {journal} {\bibinfo
			{journal} {Z. Phys. B Con. Mat.}\ }\textbf {\bibinfo {volume} {66}},\
		\bibinfo {pages} {153} (\bibinfo {year} {1987})}\BibitemShut {NoStop}%
	\bibitem [{\citenamefont {Cowan}(1981)}]{CowanBook}%
	\BibitemOpen
	\bibfield  {author} {\bibinfo {author} {\bibfnamefont {R.}~\bibnamefont
			{Cowan}},\ }\href@noop {} {\emph {\bibinfo {title} {The Theory of Atomic
				Structure and Spectra.}}}\ (\bibinfo  {publisher} {University of California,
		Berkeley},\ \bibinfo {year} {1981})\BibitemShut {NoStop}%
	\bibitem [{\citenamefont {Tanaka}\ and\ \citenamefont {Jo}(1994)}]{Tanaka1994}%
	\BibitemOpen
	\bibfield  {author} {\bibinfo {author} {\bibfnamefont {A.}~\bibnamefont
			{Tanaka}}\ and\ \bibinfo {author} {\bibfnamefont {T.}~\bibnamefont {Jo}},\
	}\bibfield  {title} {\bibinfo {title} {Resonant 3d, 3p and 3s photoemission
			in transition metal oxides predicted at 2p threshold},\ }\href
	{https://doi.org/10.1143/JPSJ.63.2788} {\bibfield  {journal} {\bibinfo
			{journal} {J Phys. Soc. Jpn.}\ }\textbf {\bibinfo {volume} {63}},\ \bibinfo
		{pages} {2788} (\bibinfo {year} {1994})}\BibitemShut {NoStop}%
	\bibitem [{\citenamefont {de~Groot}\ and\ \citenamefont
		{Kotani}(2008)}]{Groot2008}%
	\BibitemOpen
	\bibfield  {author} {\bibinfo {author} {\bibfnamefont {F.}~\bibnamefont
			{de~Groot}}\ and\ \bibinfo {author} {\bibfnamefont {A.}~\bibnamefont
			{Kotani}},\ }\href@noop {} {\emph {\bibinfo {title} {Core {L}evel
				{S}pectrocopy of {S}olis}}},\ Vol.~\bibinfo {volume} {6}\ (\bibinfo
	{publisher} {CRC Press},\ \bibinfo {year} {2008})\ \bibinfo {note}
	{{A}dvances in {C}ondensed {M}atter {S}cience}\BibitemShut {NoStop}%
	\bibitem [{\citenamefont {Strigari}\ \emph {et~al.}(2015)\citenamefont
		{Strigari}, \citenamefont {Sundermann}, \citenamefont {Muro}, \citenamefont
		{Yutani}, \citenamefont {Takabatake}, \citenamefont {Tsuei}, \citenamefont
		{Liao}, \citenamefont {Tanaka}, \citenamefont {Thalmeier}, \citenamefont
		{Haverkort}, \citenamefont {Tjeng},\ and\ \citenamefont
		{Severing}}]{Strigari2015}%
	\BibitemOpen
	\bibfield  {author} {\bibinfo {author} {\bibfnamefont {F.}~\bibnamefont
			{Strigari}}, \bibinfo {author} {\bibfnamefont {M.}~\bibnamefont
			{Sundermann}}, \bibinfo {author} {\bibfnamefont {Y.}~\bibnamefont {Muro}},
		\bibinfo {author} {\bibfnamefont {K.}~\bibnamefont {Yutani}}, \bibinfo
		{author} {\bibfnamefont {T.}~\bibnamefont {Takabatake}}, \bibinfo {author}
		{\bibfnamefont {K.-D.}\ \bibnamefont {Tsuei}}, \bibinfo {author}
		{\bibfnamefont {Y.}~\bibnamefont {Liao}}, \bibinfo {author} {\bibfnamefont
			{A.}~\bibnamefont {Tanaka}}, \bibinfo {author} {\bibfnamefont
			{P.}~\bibnamefont {Thalmeier}}, \bibinfo {author} {\bibfnamefont
			{M.}~\bibnamefont {Haverkort}}, \bibinfo {author} {\bibfnamefont
			{L.}~\bibnamefont {Tjeng}},\ and\ \bibinfo {author} {\bibfnamefont
			{A.}~\bibnamefont {Severing}},\ }\bibfield  {title} {\bibinfo {title}
		{Quantitative study of valence and configuration interaction parameters of
			the {K}ondo semiconductors {Ce$M_2$Al$_{10}$ ($M$=Ru, Os and Fe)} by means of
			bulk-sensitive hard x-ray photoelectron spectroscopy},\ }\href
	{https://doi.org/https://doi.org/10.1016/j.elspec.2015.01.004} {\bibfield
		{journal} {\bibinfo  {journal} {J. Elect. Spect. Rel. Phenom.}\ }\textbf
		{\bibinfo {volume} {199}},\ \bibinfo {pages} {56} (\bibinfo {year}
		{2015})}\BibitemShut {NoStop}%
	\bibitem [{\citenamefont {Gunnarsson}\ and\ \citenamefont
		{Sch\"onhammer}(1983)}]{Gunnarsson1983}%
	\BibitemOpen
	\bibfield  {author} {\bibinfo {author} {\bibfnamefont {O.}~\bibnamefont
			{Gunnarsson}}\ and\ \bibinfo {author} {\bibfnamefont {K.}~\bibnamefont
			{Sch\"onhammer}},\ }\bibfield  {title} {\bibinfo {title} {Electron
			spectroscopies for {C}e compounds in the impurity model},\ }\href
	{https://doi.org/10.1103/PhysRevB.28.4315} {\bibfield  {journal} {\bibinfo
			{journal} {Phys. Rev. B}\ }\textbf {\bibinfo {volume} {28}},\ \bibinfo
		{pages} {4315} (\bibinfo {year} {1983})}\BibitemShut {NoStop}%
	\bibitem [{\citenamefont {Hansmann}\ \emph {et~al.}(2008)\citenamefont
		{Hansmann}, \citenamefont {Severing}, \citenamefont {Hu}, \citenamefont
		{Haverkort}, \citenamefont {Chang}, \citenamefont {Klein}, \citenamefont
		{Tanaka}, \citenamefont {Hsieh}, \citenamefont {Lin}, \citenamefont {Chen},
		\citenamefont {F\aa{}k}, \citenamefont {Lejay},\ and\ \citenamefont
		{Tjeng}}]{Hansmann2008}%
	\BibitemOpen
	\bibfield  {author} {\bibinfo {author} {\bibfnamefont {P.}~\bibnamefont
			{Hansmann}}, \bibinfo {author} {\bibfnamefont {A.}~\bibnamefont {Severing}},
		\bibinfo {author} {\bibfnamefont {Z.}~\bibnamefont {Hu}}, \bibinfo {author}
		{\bibfnamefont {M.~W.}\ \bibnamefont {Haverkort}}, \bibinfo {author}
		{\bibfnamefont {C.~F.}\ \bibnamefont {Chang}}, \bibinfo {author}
		{\bibfnamefont {S.}~\bibnamefont {Klein}}, \bibinfo {author} {\bibfnamefont
			{A.}~\bibnamefont {Tanaka}}, \bibinfo {author} {\bibfnamefont {H.~H.}\
			\bibnamefont {Hsieh}}, \bibinfo {author} {\bibfnamefont {H.-J.}\ \bibnamefont
			{Lin}}, \bibinfo {author} {\bibfnamefont {C.~T.}\ \bibnamefont {Chen}},
		\bibinfo {author} {\bibfnamefont {B.}~\bibnamefont {F\aa{}k}}, \bibinfo
		{author} {\bibfnamefont {P.}~\bibnamefont {Lejay}},\ and\ \bibinfo {author}
		{\bibfnamefont {L.~H.}\ \bibnamefont {Tjeng}},\ }\bibfield  {title} {\bibinfo
		{title} {Determining the crystal-field ground state in rare earth heavy
			fermion materials using soft-x-ray absorption spectroscopy},\ }\href
	{https://doi.org/10.1103/PhysRevLett.100.066405} {\bibfield  {journal}
		{\bibinfo  {journal} {Phys. Rev. Lett.}\ }\textbf {\bibinfo {volume} {100}},\
		\bibinfo {pages} {066405} (\bibinfo {year} {2008})}\BibitemShut {NoStop}%
	\bibitem [{\citenamefont {Sundermann}\ \emph {et~al.}(2015)\citenamefont
		{Sundermann}, \citenamefont {Strigari}, \citenamefont {Willers},
		\citenamefont {Winkler}, \citenamefont {Prokofiev}, \citenamefont {Ablett},
		\citenamefont {Rueff}, \citenamefont {Schmitz}, \citenamefont {Weschke},
		\citenamefont {Sala}, \citenamefont {Al-Zein}, \citenamefont {Tanaka},
		\citenamefont {Haverkort}, \citenamefont {Kasinathan}, \citenamefont {Tjeng},
		\citenamefont {Paschen},\ and\ \citenamefont {Severing}}]{Sundermann2015}%
	\BibitemOpen
	\bibfield  {author} {\bibinfo {author} {\bibfnamefont {M.}~\bibnamefont
			{Sundermann}}, \bibinfo {author} {\bibfnamefont {F.}~\bibnamefont
			{Strigari}}, \bibinfo {author} {\bibfnamefont {T.}~\bibnamefont {Willers}},
		\bibinfo {author} {\bibfnamefont {H.}~\bibnamefont {Winkler}}, \bibinfo
		{author} {\bibfnamefont {A.}~\bibnamefont {Prokofiev}}, \bibinfo {author}
		{\bibfnamefont {J.~M.}\ \bibnamefont {Ablett}}, \bibinfo {author}
		{\bibfnamefont {J.-P.}\ \bibnamefont {Rueff}}, \bibinfo {author}
		{\bibfnamefont {D.}~\bibnamefont {Schmitz}}, \bibinfo {author} {\bibfnamefont
			{E.}~\bibnamefont {Weschke}}, \bibinfo {author} {\bibfnamefont {M.~M.}\
			\bibnamefont {Sala}}, \bibinfo {author} {\bibfnamefont {A.}~\bibnamefont
			{Al-Zein}}, \bibinfo {author} {\bibfnamefont {A.}~\bibnamefont {Tanaka}},
		\bibinfo {author} {\bibfnamefont {M.~W.}\ \bibnamefont {Haverkort}}, \bibinfo
		{author} {\bibfnamefont {D.}~\bibnamefont {Kasinathan}}, \bibinfo {author}
		{\bibfnamefont {L.~H.}\ \bibnamefont {Tjeng}}, \bibinfo {author}
		{\bibfnamefont {S.}~\bibnamefont {Paschen}},\ and\ \bibinfo {author}
		{\bibfnamefont {A.}~\bibnamefont {Severing}},\ }\bibfield  {title} {\bibinfo
		{title} {{CeRu$_4$Sn$_6$}: a strongly correlated material with nontrivial
			topology},\ }\href {https://doi.org/10.1038/srep17937} {\bibfield  {journal}
		{\bibinfo  {journal} {Sci. Rep.}\ }\textbf {\bibinfo {volume} {5}},\ \bibinfo
		{pages} {17937} (\bibinfo {year} {2015})}\BibitemShut {NoStop}%
	\bibitem [{\citenamefont {Wissgott}\ and\ \citenamefont
		{Held}(2016)}]{Wissgott2016}%
	\BibitemOpen
	\bibfield  {author} {\bibinfo {author} {\bibfnamefont {P.}~\bibnamefont
			{Wissgott}}\ and\ \bibinfo {author} {\bibfnamefont {K.}~\bibnamefont
			{Held}},\ }\bibfield  {title} {\bibinfo {title} {Electronic structure of
			{CeRu$_4$Sn$_6$}: a density functional plus dynamical mean field theory
			study},\ }\bibfield  {journal} {\bibinfo  {journal} {Eur. Phys. J. B}\
	}\textbf {\bibinfo {volume} {89}},\ \href {https://doi.org/10} {10} (\bibinfo
	{year} {2016})\BibitemShut {NoStop}%
	\bibitem [{\citenamefont {Amorese}\ \emph {et~al.}(2023)\citenamefont
		{Amorese}, \citenamefont {Hansmann}, \citenamefont {Marino}, \citenamefont
		{K\"orner}, \citenamefont {Willers}, \citenamefont {Walters}, \citenamefont
		{Zhou}, \citenamefont {Kummer}, \citenamefont {Brookes}, \citenamefont {Lin},
		\citenamefont {Chen}, \citenamefont {Lejay}, \citenamefont {Haverkort},
		\citenamefont {Tjeng},\ and\ \citenamefont {Severing}}]{Amorese2023}%
	\BibitemOpen
	\bibfield  {author} {\bibinfo {author} {\bibfnamefont {A.}~\bibnamefont
			{Amorese}}, \bibinfo {author} {\bibfnamefont {P.}~\bibnamefont {Hansmann}},
		\bibinfo {author} {\bibfnamefont {A.}~\bibnamefont {Marino}}, \bibinfo
		{author} {\bibfnamefont {P.}~\bibnamefont {K\"orner}}, \bibinfo {author}
		{\bibfnamefont {T.}~\bibnamefont {Willers}}, \bibinfo {author} {\bibfnamefont
			{A.}~\bibnamefont {Walters}}, \bibinfo {author} {\bibfnamefont {K.-J.}\
			\bibnamefont {Zhou}}, \bibinfo {author} {\bibfnamefont {K.}~\bibnamefont
			{Kummer}}, \bibinfo {author} {\bibfnamefont {N.~B.}\ \bibnamefont {Brookes}},
		\bibinfo {author} {\bibfnamefont {H.-J.}\ \bibnamefont {Lin}}, \bibinfo
		{author} {\bibfnamefont {C.-T.}\ \bibnamefont {Chen}}, \bibinfo {author}
		{\bibfnamefont {P.}~\bibnamefont {Lejay}}, \bibinfo {author} {\bibfnamefont
			{M.~W.}\ \bibnamefont {Haverkort}}, \bibinfo {author} {\bibfnamefont {L.~H.}\
			\bibnamefont {Tjeng}},\ and\ \bibinfo {author} {\bibfnamefont
			{A.}~\bibnamefont {Severing}},\ }\bibfield  {title} {\bibinfo {title}
		{Orbital selective coupling in {CeRh}$_3${B}$_2$: Coexistence of high {C}urie
			and high {K}ondo temperatures},\ }\href
	{https://doi.org/10.1103/PhysRevB.107.115164} {\bibfield  {journal} {\bibinfo
			{journal} {Phys. Rev. B}\ }\textbf {\bibinfo {volume} {107}},\ \bibinfo
		{pages} {115164} (\bibinfo {year} {2023})}\BibitemShut {NoStop}%
	\bibitem [{\citenamefont {Christovam}\ \emph {et~al.}(2024)\citenamefont
		{Christovam}, \citenamefont {Ferreira-Carvalho}, \citenamefont {Marino},
		\citenamefont {Sundermann}, \citenamefont {Takegami}, \citenamefont
		{Melendez-Sans}, \citenamefont {Tsuei}, \citenamefont {Hu}, \citenamefont
		{R\"o\ss{}ler}, \citenamefont {Valvidares}, \citenamefont {Haverkort},
		\citenamefont {Liu}, \citenamefont {Bauer}, \citenamefont {Tjeng},
		\citenamefont {Zwicknagl},\ and\ \citenamefont {Severing}}]{Christovam2024}%
	\BibitemOpen
	\bibfield  {author} {\bibinfo {author} {\bibfnamefont {D.~S.}\ \bibnamefont
			{Christovam}}, \bibinfo {author} {\bibfnamefont {M.}~\bibnamefont
			{Ferreira-Carvalho}}, \bibinfo {author} {\bibfnamefont {A.}~\bibnamefont
			{Marino}}, \bibinfo {author} {\bibfnamefont {M.}~\bibnamefont {Sundermann}},
		\bibinfo {author} {\bibfnamefont {D.}~\bibnamefont {Takegami}}, \bibinfo
		{author} {\bibfnamefont {A.}~\bibnamefont {Melendez-Sans}}, \bibinfo {author}
		{\bibfnamefont {K.~D.}\ \bibnamefont {Tsuei}}, \bibinfo {author}
		{\bibfnamefont {Z.}~\bibnamefont {Hu}}, \bibinfo {author} {\bibfnamefont
			{S.}~\bibnamefont {R\"o\ss{}ler}}, \bibinfo {author} {\bibfnamefont
			{M.}~\bibnamefont {Valvidares}}, \bibinfo {author} {\bibfnamefont {M.~W.}\
			\bibnamefont {Haverkort}}, \bibinfo {author} {\bibfnamefont {Y.}~\bibnamefont
			{Liu}}, \bibinfo {author} {\bibfnamefont {E.~D.}\ \bibnamefont {Bauer}},
		\bibinfo {author} {\bibfnamefont {L.~H.}\ \bibnamefont {Tjeng}}, \bibinfo
		{author} {\bibfnamefont {G.}~\bibnamefont {Zwicknagl}},\ and\ \bibinfo
		{author} {\bibfnamefont {A.}~\bibnamefont {Severing}},\ }\bibfield  {title}
	{\bibinfo {title} {Spectroscopic evidence of {K}ondo-induced quasiquartet in
			{CeRh}$_2${As}$_2$},\ }\href {https://doi.org/10.1103/PhysRevLett.132.046401}
	{\bibfield  {journal} {\bibinfo  {journal} {Phys. Rev. Lett.}\ }\textbf
		{\bibinfo {volume} {132}},\ \bibinfo {pages} {046401} (\bibinfo {year}
		{2024})}\BibitemShut {NoStop}%
	\bibitem [{\citenamefont {Sundermann}\ \emph {et~al.}(2016)\citenamefont
		{Sundermann}, \citenamefont {Strigari}, \citenamefont {Willers},
		\citenamefont {Weinen}, \citenamefont {Liao}, \citenamefont {Tsuei},
		\citenamefont {Hiraoka}, \citenamefont {Ishii}, \citenamefont {Yamaoka},
		\citenamefont {Mizuki}, \citenamefont {Zekko}, \citenamefont {Bauer},
		\citenamefont {Sarrao}, \citenamefont {Thompson}, \citenamefont {Lejay},
		\citenamefont {Muro}, \citenamefont {Yutani}, \citenamefont {Takabatake},
		\citenamefont {Tanaka}, \citenamefont {Hollmann}, \citenamefont {Tjeng},\
		and\ \citenamefont {Severing}}]{Sundermann2016}%
	\BibitemOpen
	\bibfield  {author} {\bibinfo {author} {\bibfnamefont {M.}~\bibnamefont
			{Sundermann}}, \bibinfo {author} {\bibfnamefont {F.}~\bibnamefont
			{Strigari}}, \bibinfo {author} {\bibfnamefont {T.}~\bibnamefont {Willers}},
		\bibinfo {author} {\bibfnamefont {J.}~\bibnamefont {Weinen}}, \bibinfo
		{author} {\bibfnamefont {Y.}~\bibnamefont {Liao}}, \bibinfo {author}
		{\bibfnamefont {K.-D.}\ \bibnamefont {Tsuei}}, \bibinfo {author}
		{\bibfnamefont {N.}~\bibnamefont {Hiraoka}}, \bibinfo {author} {\bibfnamefont
			{H.}~\bibnamefont {Ishii}}, \bibinfo {author} {\bibfnamefont
			{H.}~\bibnamefont {Yamaoka}}, \bibinfo {author} {\bibfnamefont
			{J.}~\bibnamefont {Mizuki}}, \bibinfo {author} {\bibfnamefont
			{Y.}~\bibnamefont {Zekko}}, \bibinfo {author} {\bibfnamefont
			{E.}~\bibnamefont {Bauer}}, \bibinfo {author} {\bibfnamefont
			{J.}~\bibnamefont {Sarrao}}, \bibinfo {author} {\bibfnamefont
			{J.}~\bibnamefont {Thompson}}, \bibinfo {author} {\bibfnamefont
			{P.}~\bibnamefont {Lejay}}, \bibinfo {author} {\bibfnamefont
			{Y.}~\bibnamefont {Muro}}, \bibinfo {author} {\bibfnamefont {K.}~\bibnamefont
			{Yutani}}, \bibinfo {author} {\bibfnamefont {T.}~\bibnamefont {Takabatake}},
		\bibinfo {author} {\bibfnamefont {A.}~\bibnamefont {Tanaka}}, \bibinfo
		{author} {\bibfnamefont {N.}~\bibnamefont {Hollmann}}, \bibinfo {author}
		{\bibfnamefont {L.}~\bibnamefont {Tjeng}},\ and\ \bibinfo {author}
		{\bibfnamefont {A.}~\bibnamefont {Severing}},\ }\bibfield  {title} {\bibinfo
		{title} {Quantitative study of the $f$ occupation in {Ce$M$In$_5$} and other
			cerium compounds with hard x-rays},\ }\href
	{https://doi.org/https://doi.org/10.1016/j.elspec.2016.02.002} {\bibfield
		{journal} {\bibinfo  {journal} {J. Elec. Spect. and Rel. Phen.}\ }\textbf
		{\bibinfo {volume} {209}},\ \bibinfo {pages} {1} (\bibinfo {year}
		{2016})}\BibitemShut {NoStop}%
	\bibitem [{\citenamefont {Dhar}\ \emph {et~al.}(1981)\citenamefont {Dhar},
		\citenamefont {Malik},\ and\ \citenamefont {Vijayaraghavan}}]{Dhar1981}%
	\BibitemOpen
	\bibfield  {author} {\bibinfo {author} {\bibfnamefont {S.~K.}\ \bibnamefont
			{Dhar}}, \bibinfo {author} {\bibfnamefont {S.~K.}\ \bibnamefont {Malik}},\
		and\ \bibinfo {author} {\bibfnamefont {R.}~\bibnamefont {Vijayaraghavan}},\
	}\bibfield  {title} {\bibinfo {title} {Strong itinerant magnetism in ternary
			boride {CeRh$_3$B$_2$}},\ }\href
	{https://doi.org/10.1088/0022-3719/14/11/008} {\bibfield  {journal} {\bibinfo
			{journal} {J. Phys. C: Solid State Phys.}\ }\textbf {\bibinfo {volume}
			{14}},\ \bibinfo {pages} {L321} (\bibinfo {year} {1981})}\BibitemShut
	{NoStop}%
	\bibitem [{\citenamefont {Galatanu}\ \emph {et~al.}(2003)\citenamefont
		{Galatanu}, \citenamefont {Yamamoto}, \citenamefont {Okubo}, \citenamefont
		{Yamada}, \citenamefont {Thamizhavel}, \citenamefont {Takeuchi},
		\citenamefont {Sugiyama}, \citenamefont {Inada},\ and\ \citenamefont
		{Onuki}}]{Galatanu2003}%
	\BibitemOpen
	\bibfield  {author} {\bibinfo {author} {\bibfnamefont {A.}~\bibnamefont
			{Galatanu}}, \bibinfo {author} {\bibfnamefont {E.}~\bibnamefont {Yamamoto}},
		\bibinfo {author} {\bibfnamefont {T.}~\bibnamefont {Okubo}}, \bibinfo
		{author} {\bibfnamefont {M.}~\bibnamefont {Yamada}}, \bibinfo {author}
		{\bibfnamefont {A.}~\bibnamefont {Thamizhavel}}, \bibinfo {author}
		{\bibfnamefont {T.}~\bibnamefont {Takeuchi}}, \bibinfo {author}
		{\bibfnamefont {K.}~\bibnamefont {Sugiyama}}, \bibinfo {author}
		{\bibfnamefont {Y.}~\bibnamefont {Inada}},\ and\ \bibinfo {author}
		{\bibfnamefont {Y.}~\bibnamefont {Onuki}},\ }\bibfield  {title} {\bibinfo
		{title} {On the unusual magnetic behaviour of {CeRh$_3$B$_2$}},\ }\href
	{https://doi.org/10.1088/0953-8984/15/28/349} {\bibfield  {journal} {\bibinfo
			{journal} {J. Phys.: Cond. Mat.}\ }\textbf {\bibinfo {volume} {15}},\
		\bibinfo {pages} {S2187} (\bibinfo {year} {2003})}\BibitemShut {NoStop}%
	\bibitem [{\citenamefont {Malik}\ \emph {et~al.}(1983)\citenamefont {Malik},
		\citenamefont {Vijayaraghavan}, \citenamefont {Wallace},\ and\ \citenamefont
		{Dhar}}]{Malik1983}%
	\BibitemOpen
	\bibfield  {author} {\bibinfo {author} {\bibfnamefont {S.}~\bibnamefont
			{Malik}}, \bibinfo {author} {\bibfnamefont {R.}~\bibnamefont
			{Vijayaraghavan}}, \bibinfo {author} {\bibfnamefont {W.}~\bibnamefont
			{Wallace}},\ and\ \bibinfo {author} {\bibfnamefont {S.}~\bibnamefont
			{Dhar}},\ }\bibfield  {title} {\bibinfo {title} {Magnetic behavior of
			{R$_3$B$_2$} ({R = La to Gd}) ternary borides},\ }\href
	{https://doi.org/https://doi.org/10.1016/0304-8853(83)90060-4} {\bibfield
		{journal} {\bibinfo  {journal} {J. Mag. Mag. Mat.}\ }\textbf {\bibinfo
			{volume} {37}},\ \bibinfo {pages} {303 } (\bibinfo {year}
		{1983})}\BibitemShut {NoStop}%
	\bibitem [{\citenamefont {Maple}\ \emph {et~al.}(1985)\citenamefont {Maple},
		\citenamefont {Lambert}, \citenamefont {Torikachvili}, \citenamefont {Yang},
		\citenamefont {Allen}, \citenamefont {Pate},\ and\ \citenamefont
		{Lindau}}]{Maple1985}%
	\BibitemOpen
	\bibfield  {author} {\bibinfo {author} {\bibfnamefont {M.}~\bibnamefont
			{Maple}}, \bibinfo {author} {\bibfnamefont {S.}~\bibnamefont {Lambert}},
		\bibinfo {author} {\bibfnamefont {M.}~\bibnamefont {Torikachvili}}, \bibinfo
		{author} {\bibfnamefont {K.}~\bibnamefont {Yang}}, \bibinfo {author}
		{\bibfnamefont {J.}~\bibnamefont {Allen}}, \bibinfo {author} {\bibfnamefont
			{B.}~\bibnamefont {Pate}},\ and\ \bibinfo {author} {\bibfnamefont
			{I.}~\bibnamefont {Lindau}},\ }\bibfield  {title} {\bibinfo {title}
		{Superconductivity, magnetism and valence fluctuations in rare
			earth-transition metal borides},\ }\href
	{https://doi.org/https://doi.org/10.1016/0022-5088(85)90192-4} {\bibfield
		{journal} {\bibinfo  {journal} {J. Less Common Met.}\ }\textbf {\bibinfo
			{volume} {111}},\ \bibinfo {pages} {239 } (\bibinfo {year}
		{1985})}\BibitemShut {NoStop}%
	\bibitem [{\citenamefont {Kitaoka}\ \emph {et~al.}(1985)\citenamefont
		{Kitaoka}, \citenamefont {Kishimoto}, \citenamefont {Asayama}, \citenamefont
		{Kohara}, \citenamefont {Takeda}, \citenamefont {Vijayaraghavan},
		\citenamefont {Malik}, \citenamefont {Dhar},\ and\ \citenamefont
		{Rambabu}}]{Kitaoka1985}%
	\BibitemOpen
	\bibfield  {author} {\bibinfo {author} {\bibfnamefont {Y.}~\bibnamefont
			{Kitaoka}}, \bibinfo {author} {\bibfnamefont {Y.}~\bibnamefont {Kishimoto}},
		\bibinfo {author} {\bibfnamefont {K.}~\bibnamefont {Asayama}}, \bibinfo
		{author} {\bibfnamefont {T.}~\bibnamefont {Kohara}}, \bibinfo {author}
		{\bibfnamefont {K.}~\bibnamefont {Takeda}}, \bibinfo {author} {\bibfnamefont
			{R.}~\bibnamefont {Vijayaraghavan}}, \bibinfo {author} {\bibfnamefont
			{S.}~\bibnamefont {Malik}}, \bibinfo {author} {\bibfnamefont
			{S.}~\bibnamefont {Dhar}},\ and\ \bibinfo {author} {\bibfnamefont
			{D.}~\bibnamefont {Rambabu}},\ }\bibfield  {title} {\bibinfo {title}
		{Magnetic behavior of {RERh$_3$B$_2$} ({RE = La, Ce, Nd and Gd})},\ }\href
	{https://doi.org/https://doi.org/10.1016/0304-8853(85)90330-0} {\bibfield
		{journal} {\bibinfo  {journal} {J. Mag. Mag. Mat.}\ }\textbf {\bibinfo
			{volume} {52}},\ \bibinfo {pages} {449 } (\bibinfo {year}
		{1985})}\BibitemShut {NoStop}%
	\bibitem [{\citenamefont {Sampathkumaran}\ \emph {et~al.}(1985)\citenamefont
		{Sampathkumaran}, \citenamefont {Kaindl}, \citenamefont {Laubschat},
		\citenamefont {Krone},\ and\ \citenamefont {Wortmann}}]{Sampa1985}%
	\BibitemOpen
	\bibfield  {author} {\bibinfo {author} {\bibfnamefont {E.~V.}\ \bibnamefont
			{Sampathkumaran}}, \bibinfo {author} {\bibfnamefont {G.}~\bibnamefont
			{Kaindl}}, \bibinfo {author} {\bibfnamefont {C.}~\bibnamefont {Laubschat}},
		\bibinfo {author} {\bibfnamefont {W.}~\bibnamefont {Krone}},\ and\ \bibinfo
		{author} {\bibfnamefont {G.}~\bibnamefont {Wortmann}},\ }\bibfield  {title}
	{\bibinfo {title} {Spectroscopic evidence against rh 4d itinerant
			ferromagnetism in {CeRh$_3$B$_2$}},\ }\href
	{https://doi.org/10.1103/PhysRevB.31.3185} {\bibfield  {journal} {\bibinfo
			{journal} {Phys. Rev. B}\ }\textbf {\bibinfo {volume} {31}},\ \bibinfo
		{pages} {3185} (\bibinfo {year} {1985})}\BibitemShut {NoStop}%
	\bibitem [{\citenamefont {Shaheen}\ \emph {et~al.}(1985)\citenamefont
		{Shaheen}, \citenamefont {Schilling},\ and\ \citenamefont
		{Shelton}}]{Shaheen1985}%
	\BibitemOpen
	\bibfield  {author} {\bibinfo {author} {\bibfnamefont {S.~A.}\ \bibnamefont
			{Shaheen}}, \bibinfo {author} {\bibfnamefont {J.~S.}\ \bibnamefont
			{Schilling}},\ and\ \bibinfo {author} {\bibfnamefont {R.~N.}\ \bibnamefont
			{Shelton}},\ }\bibfield  {title} {\bibinfo {title} {Anomalous ferromagnetism
			in {CeRh}$_3${B}$_2$: Possibility of a new {K}ondo-lattice state},\ }\href
	{https://doi.org/10.1103/PhysRevB.31.656} {\bibfield  {journal} {\bibinfo
			{journal} {Phys. Rev. B}\ }\textbf {\bibinfo {volume} {31}},\ \bibinfo
		{pages} {656} (\bibinfo {year} {1985})}\BibitemShut {NoStop}%
	\bibitem [{\citenamefont {Fujimori}\ \emph {et~al.}(1990)\citenamefont
		{Fujimori}, \citenamefont {Takahashi}, \citenamefont {Okabe}, \citenamefont
		{Kasaya},\ and\ \citenamefont {Kasuya}}]{Fujimori1990}%
	\BibitemOpen
	\bibfield  {author} {\bibinfo {author} {\bibfnamefont {A.}~\bibnamefont
			{Fujimori}}, \bibinfo {author} {\bibfnamefont {T.}~\bibnamefont {Takahashi}},
		\bibinfo {author} {\bibfnamefont {A.}~\bibnamefont {Okabe}}, \bibinfo
		{author} {\bibfnamefont {M.}~\bibnamefont {Kasaya}},\ and\ \bibinfo {author}
		{\bibfnamefont {T.}~\bibnamefont {Kasuya}},\ }\bibfield  {title} {\bibinfo
		{title} {Photoemission study of the ferromagnetic {K}ondo system
			{CeRh$_3$B$_2$}},\ }\href {https://doi.org/10.1103/PhysRevB.41.6783}
	{\bibfield  {journal} {\bibinfo  {journal} {Phys. Rev. B}\ }\textbf {\bibinfo
			{volume} {41}},\ \bibinfo {pages} {6783} (\bibinfo {year}
		{1990})}\BibitemShut {NoStop}%
	\bibitem [{\citenamefont {Allen}\ \emph {et~al.}(1990)\citenamefont {Allen},
		\citenamefont {Maple}, \citenamefont {Kang}, \citenamefont {Yang},
		\citenamefont {Torikachvili}, \citenamefont {Lassailly}, \citenamefont
		{Ellis}, \citenamefont {Pate},\ and\ \citenamefont {Lindau}}]{Allen1990}%
	\BibitemOpen
	\bibfield  {author} {\bibinfo {author} {\bibfnamefont {J.~W.}\ \bibnamefont
			{Allen}}, \bibinfo {author} {\bibfnamefont {M.~B.}\ \bibnamefont {Maple}},
		\bibinfo {author} {\bibfnamefont {J.-S.}\ \bibnamefont {Kang}}, \bibinfo
		{author} {\bibfnamefont {K.~N.}\ \bibnamefont {Yang}}, \bibinfo {author}
		{\bibfnamefont {M.~S.}\ \bibnamefont {Torikachvili}}, \bibinfo {author}
		{\bibfnamefont {Y.}~\bibnamefont {Lassailly}}, \bibinfo {author}
		{\bibfnamefont {W.~P.}\ \bibnamefont {Ellis}}, \bibinfo {author}
		{\bibfnamefont {B.~B.}\ \bibnamefont {Pate}},\ and\ \bibinfo {author}
		{\bibfnamefont {I.}~\bibnamefont {Lindau}},\ }\bibfield  {title} {\bibinfo
		{title} {Density-of-states-driven transition from superconductivity to
			ferromagnetism in {CeRh$_3$B$_2$}: Scenario for an exchange-split {K}ondo
			resonance},\ }\href {https://doi.org/10.1103/PhysRevB.41.9013} {\bibfield
		{journal} {\bibinfo  {journal} {Phys. Rev. B}\ }\textbf {\bibinfo {volume}
			{41}},\ \bibinfo {pages} {9013} (\bibinfo {year} {1990})}\BibitemShut
	{NoStop}%
	\bibitem [{\citenamefont {Takeuchi}\ \emph {et~al.}(2004)\citenamefont
		{Takeuchi}, \citenamefont {Thamizhavel}, \citenamefont {Okubo}, \citenamefont
		{Yamada}, \citenamefont {Inada}, \citenamefont {Galatanu}, \citenamefont
		{Yamamoto},\ and\ \citenamefont {Ōnuki}}]{Takeuchi2004}%
	\BibitemOpen
	\bibfield  {author} {\bibinfo {author} {\bibfnamefont {T.}~\bibnamefont
			{Takeuchi}}, \bibinfo {author} {\bibfnamefont {A.}~\bibnamefont
			{Thamizhavel}}, \bibinfo {author} {\bibfnamefont {T.}~\bibnamefont {Okubo}},
		\bibinfo {author} {\bibfnamefont {M.}~\bibnamefont {Yamada}}, \bibinfo
		{author} {\bibfnamefont {Y.}~\bibnamefont {Inada}}, \bibinfo {author}
		{\bibfnamefont {A.}~\bibnamefont {Galatanu}}, \bibinfo {author}
		{\bibfnamefont {E.}~\bibnamefont {Yamamoto}},\ and\ \bibinfo {author}
		{\bibfnamefont {Y.}~\bibnamefont {Ōnuki}},\ }\bibfield  {title} {\bibinfo
		{title} {Thermal expansion and magnetostriction in {CeRh$_3$B$_2$}},\ }\href
	{https://doi.org/https://doi.org/10.1016/j.jmmm.2003.12.1307} {\bibfield
		{journal} {\bibinfo  {journal} {J. Mag. Mag. Mat.}\ }\textbf {\bibinfo
			{volume} {272-276}},\ \bibinfo {pages} {E17 } (\bibinfo {year} {2004})},\
	\bibinfo {note} {proceedings of the International Conference on Magnetism
		(ICM 2003)}\BibitemShut {NoStop}%
	\bibitem [{\citenamefont {Zevin}\ \emph {et~al.}(1988)\citenamefont {Zevin},
		\citenamefont {Zwicknagl},\ and\ \citenamefont {Fulde}}]{Zevin1988}%
	\BibitemOpen
	\bibfield  {author} {\bibinfo {author} {\bibfnamefont {V.}~\bibnamefont
			{Zevin}}, \bibinfo {author} {\bibfnamefont {G.}~\bibnamefont {Zwicknagl}},\
		and\ \bibinfo {author} {\bibfnamefont {P.}~\bibnamefont {Fulde}},\ }\bibfield
	{title} {\bibinfo {title} {Temperature dependence of the $4f$ quadrupole
			moment of {Y}b in {YbCu}$2${Si}$_2$},\ }\href
	{https://doi.org/10.1103/PhysRevLett.60.2331} {\bibfield  {journal} {\bibinfo
			{journal} {Phys. Rev. Lett.}\ }\textbf {\bibinfo {volume} {60}},\ \bibinfo
		{pages} {2331} (\bibinfo {year} {1988})}\BibitemShut {NoStop}%
	\bibitem [{\citenamefont {Zwicknagl}\ \emph {et~al.}(1990)\citenamefont
		{Zwicknagl}, \citenamefont {Zevin},\ and\ \citenamefont
		{Fulde}}]{Zwicknagl1990}%
	\BibitemOpen
	\bibfield  {author} {\bibinfo {author} {\bibfnamefont {G.}~\bibnamefont
			{Zwicknagl}}, \bibinfo {author} {\bibfnamefont {V.}~\bibnamefont {Zevin}},\
		and\ \bibinfo {author} {\bibfnamefont {P.}~\bibnamefont {Fulde}},\ }\bibfield
	{title} {\bibinfo {title} {Simple approximation scheme for the {A}nderson
			impurity {H}amiltonian},\ }\href {https://doi.org/10.1007/BF01437646}
	{\bibfield  {journal} {\bibinfo  {journal} {Z. Phys. B Cond. Matter}\
		}\textbf {\bibinfo {volume} {97}},\ \bibinfo {pages} {365} (\bibinfo {year}
		{1990})}\BibitemShut {NoStop}%
	\bibitem [{\citenamefont {Amorese}\ \emph {et~al.}(2020)\citenamefont
		{Amorese}, \citenamefont {Marino}, \citenamefont {Sundermann}, \citenamefont
		{Chen}, \citenamefont {Hu}, \citenamefont {Willers}, \citenamefont
		{Choueikani}, \citenamefont {Ohresser}, \citenamefont {Herrero-Martin},
		\citenamefont {Agrestini}, \citenamefont {Chen}, \citenamefont {Lin},
		\citenamefont {Haverkort}, \citenamefont {Seiro}, \citenamefont {Geibel},
		\citenamefont {Steglich}, \citenamefont {Tjeng}, \citenamefont {Zwicknagl},\
		and\ \citenamefont {Severing}}]{Amorese2020}%
	\BibitemOpen
	\bibfield  {author} {\bibinfo {author} {\bibfnamefont {A.}~\bibnamefont
			{Amorese}}, \bibinfo {author} {\bibfnamefont {A.}~\bibnamefont {Marino}},
		\bibinfo {author} {\bibfnamefont {M.}~\bibnamefont {Sundermann}}, \bibinfo
		{author} {\bibfnamefont {K.}~\bibnamefont {Chen}}, \bibinfo {author}
		{\bibfnamefont {Z.}~\bibnamefont {Hu}}, \bibinfo {author} {\bibfnamefont
			{T.}~\bibnamefont {Willers}}, \bibinfo {author} {\bibfnamefont
			{F.}~\bibnamefont {Choueikani}}, \bibinfo {author} {\bibfnamefont
			{P.}~\bibnamefont {Ohresser}}, \bibinfo {author} {\bibfnamefont
			{J.}~\bibnamefont {Herrero-Martin}}, \bibinfo {author} {\bibfnamefont
			{S.}~\bibnamefont {Agrestini}}, \bibinfo {author} {\bibfnamefont
			{C.}~\bibnamefont {Chen}}, \bibinfo {author} {\bibfnamefont {H.-J.}\
			\bibnamefont {Lin}}, \bibinfo {author} {\bibfnamefont {M.~W.}\ \bibnamefont
			{Haverkort}}, \bibinfo {author} {\bibfnamefont {S.}~\bibnamefont {Seiro}},
		\bibinfo {author} {\bibfnamefont {C.}~\bibnamefont {Geibel}}, \bibinfo
		{author} {\bibfnamefont {F.}~\bibnamefont {Steglich}}, \bibinfo {author}
		{\bibfnamefont {L.~H.}\ \bibnamefont {Tjeng}}, \bibinfo {author}
		{\bibfnamefont {G.}~\bibnamefont {Zwicknagl}},\ and\ \bibinfo {author}
		{\bibfnamefont {A.}~\bibnamefont {Severing}},\ }\bibfield  {title} {\bibinfo
		{title} {Possible multiorbital ground state in {CeCu}$_2${Si}$_2$},\ }\href
	{https://doi.org/10.1103/PhysRevB.102.245146} {\bibfield  {journal} {\bibinfo
			{journal} {Phys. Rev. B}\ }\textbf {\bibinfo {volume} {102}},\ \bibinfo
		{pages} {245146} (\bibinfo {year} {2020})}\BibitemShut {NoStop}%
	\bibitem [{\citenamefont {Rajan}(1983)}]{Rajan1983}%
	\BibitemOpen
	\bibfield  {author} {\bibinfo {author} {\bibfnamefont {V.~T.}\ \bibnamefont
			{Rajan}},\ }\bibfield  {title} {\bibinfo {title} {Magnetic susceptibility and
			specific heat of the {Coqblin-Schrieffer} model},\ }\href
	{https://doi.org/10.1103/PhysRevLett.51.308} {\bibfield  {journal} {\bibinfo
			{journal} {Phys. Rev. Lett.}\ }\textbf {\bibinfo {volume} {51}},\ \bibinfo
		{pages} {308} (\bibinfo {year} {1983})}\BibitemShut {NoStop}%
	\bibitem [{\citenamefont {Thompson}\ and\ \citenamefont
		{Lawrence}(1994)}]{Thompson1994}%
	\BibitemOpen
	\bibfield  {author} {\bibinfo {author} {\bibfnamefont {J.~D.}\ \bibnamefont
			{Thompson}}\ and\ \bibinfo {author} {\bibfnamefont {J.~M.}\ \bibnamefont
			{Lawrence}},\ }\bibfield  {title} {\bibinfo {title} {{in Handbook on the
				Physics and Chemistry of Rare Earths}}\ }(\bibinfo  {publisher} {Elsevier},\
	\bibinfo {address} {North Holland, Amsterdam},\ \bibinfo {year} {1994})\ p.\
	\bibinfo {pages} {383},\ \bibinfo {note} {{High Pressure Studies - Physical
			Properties of Anomalous Ce, Yb and U Compounds.}}\BibitemShut {Stop}%
\end{thebibliography}

\end{document}